\documentclass[12pt,a4paper]{article}
\usepackage[latin1]{inputenc}
\usepackage{a4wide}
\usepackage{amsfonts}
\usepackage{amssymb}
\usepackage{amsmath,cite}
\usepackage{bbm}
\usepackage{ifpdf}
\ifpdf
\usepackage[pdftex]{graphicx}
\usepackage[pdftex,unicode,implicit]{hyperref}

\hypersetup{%
  pdftitle    = {Gauge Theories, Duality Relations and the Tensor Hierarchy},
  pdfkeywords = {duality, gauge, gauged supergravity, symmetry, hierarchy, embedding tensor},
  pdfauthor   = {Eric A. Bergshoeff, Jelle Hartong, Olaf Hohm, Mechtild Hubscher, Tomas Ortin},
  pdfcreator  = {pdf\LaTeXe\ with package \flqq hyperref\frqq},
  pdfproducer = {pdf\LaTeXe\ with package \flqq hyperref\frqq},
  pdfpagemode = None,  
  pdffitwindow= true,  
  unicode     = true,
  plainpages  = true,
  colorlinks  = true,  
  citecolor   = blue,
  urlcolor    = red,
  linkcolor   = black
}
\newcommand{\hepth}[1]{arXiv:{\tt
\href{http://www.arXiv.org/abs/hep-th/#1}{hep-th/#1}}}

\newcommand{\arxiv}[1]{{\tt
\href{http://www.arXiv.org/abs/#1}{arXiv:#1}}}
\else
  \usepackage[dvips]{graphicx}
  \usepackage[unicode,implicit]{hyperref}
  \newcommand{\hepth}[1]{arXiv:{\tt hep-th/#1}}

  \newcommand{\arxiv}[1]{{\tt arXiv:#1}}
\fi
\makeatletter
\@addtoreset{equation}{section}
\makeatother

\makeindex
\begin{document}

\begin{flushright}
\small
UG-09-02\\
IFT-UAM/CSIC-09-01\\
January 2009\\
\normalsize
\end{flushright}

\begin{center}

\vspace{.5cm}

{\LARGE {\bf Gauge Theories, Duality Relations and the}}\\[.5cm]
{\LARGE {\bf Tensor Hierarchy}}

\vspace{.7cm}

\begin{center}

{\bf Eric A.~Bergshoeff ${}^\dagger$, Jelle Hartong ${}^\star$,
Olaf Hohm ${}^\dagger$\, \\[0.2cm]
Mechthild H\"ubscher ${}^\ddagger$, and Tom\'as Ort\'{\i}n} ${}^\ddagger$ \\[0.2cm]

\vskip 25pt

${}^\dagger$ {\em Centre for Theoretical Physics, University of Groningen, \\
Nijenborgh 4, 9747 AG Groningen, The Netherlands \vskip 5pt }

{email: {\tt E.A.Bergshoeff@rug.nl, O.Hohm@rug.nl }} \\
\vskip 15pt

${}^\star$ {\em Center for Research and Education in Fundamental Physics,\\
Institute for Theoretical Physics, \\
Sidlerstrasse 5, CH-3012 Bern,
Switzerland\vskip 5pt}

{email: {\tt hartong@itp.unibe.ch}}
\vskip 15pt

${}^\ddagger$ {\em Instituto de F\'{\i}sica Te\'orica UAM/CSIC
Facultad de Ciencias C-XVI, \\
C.U. Cantoblanco, E-28049-Madrid, Spain\vskip 5pt}

{email: {\tt Mechthild.Huebscher@uam.es, Tomas.Ortin@uam.es}}

\end{center}

\vspace{1cm}

{\bf Abstract}

\begin{quotation}

{\small
  We compute the complete 3- and 4-dimensional tensor hierarchies, i.e.~sets
  of $p$-form fields, with $1\le p\le D$, which realize an off-shell algebra
  of bosonic gauge transformations. We show how these tensor hierarchies can
  be put on-shell by introducing a set of duality relations, thereby
  introducing additional scalars and a metric tensor. These so-called duality
  hierarchies encode the equations of motion of the bosonic part of the most
  general gauged supergravity theories in those dimensions, including the
  (projected) scalar equations of motion.

  We construct gauge-invariant actions that include all the fields in the
  tensor hierarchies. We elucidate the relation between the gauge transformations of the
  $p$-form fields in the action and those of the same fields in the tensor
  hierarchy.

}

\end{quotation}

\end{center}

\newpage
\pagestyle{plain}

\tableofcontents

\newpage

\section{Introduction}

The bosonic degrees of freedom of a generic supergravity theory are
described by a metric tensor field and a set of (electric) $p$-form
potentials with $p \ge 0$. In order to describe the correct number
of degrees of freedom these fields must satisfy second-order
differential equations. In general one may realize the supersymmetry
algebra on a larger set of $p$-form potentials as long as this does
not upset the counting of degrees of freedom. Such potentials are
expected to exist in order to allow for the coupling of various
types of branes. Examples of such potentials are the (magnetic)
$(D-p-2)$-forms.  Whereas the $p$-form couples to an (electric)
$(p-1)$-brane, the $(D-p-2)$-form potential couples to a (magnetic)
$(D-p-3)$-brane. The magnetic $(D-p-2)$-forms do not describe new
degrees of freedom since they are related to the electric $p$-forms
via a first-order duality relation. By virtue of the Bianchi
identities that the curvatures of the electric and magnetic
potentials satisfy, the second-order equations can be derived as
integrability conditions of the duality relations:
\begin{equation}
\label{eq:rule} \mathrm{Bianchi\,\, identities} \;\; \& \;\;
\mathrm{duality\,\, relations} \;\Leftrightarrow\;
\mathrm{equations\,\, of\,\, motion}\, .
\end{equation}
For instance, in the case of IIA/IIB supergravity the supersymmetry algebra
can be realized on all $p$-forms ($0\le p \le 10$) with $p$ odd (IIA) or $p$
even (IIB). The Bianchi identities and duality relations then lead to all
equations of motion (except the Einstein equation). This is often referred to
as the ``democratic formulation'' of IIA/IIB supergravity
\cite{Bergshoeff:2001pv}.

The idea of deriving the equations of motion of supergravity from an
underlying set of Bianchi identities and first-order differential equations
has been pursued in several contexts in the literature. It already occurs in
the work of \cite{Cremmer:1998px} for the case of maximal supergravity
including massive IIA supergravity \cite{Lavrinenko:1999xi}. Similar duality
relations are natural in the $E_{11}$-approach to supergravity
\cite{West:2001as,Schnakenburg:2001ya,Kleinschmidt:2003mf,Riccioni:2007ni}.
Duality relations also play an important role in encoding the integrability of
a system, for instance in maximal two-dimensional supergravity
\cite{Nicolai:1987kz}.

Recently, it has been shown that dual potentials are not only relevant to
describe the coupling to branes but play also a crucial role in the
construction of a supersymmetric action for certain gauged supergravity
theories. A systematic way to study the most general gaugings of a
supergravity theory is provided by the embedding tensor approach
\cite{Nicolai:2000sc,Nicolai:2001sv,deWit:2005ub,deWit:2008ta,deWit:2008gc},
which is a powerful technique to construct in a unified way gauged
supergravity theories for different gauge groups. Usually, supersymmetric
actions involve besides the metric tensor only electric potentials. However,
using the embedding tensor approach, it has been shown that to describe a
magnetic gauging in $D=4$, i.e.~a gauging involving a magnetic vector field,
the action must also contain a dual 2-form potential via a Chern-Simons-like
topological coupling.\footnote{In the context of $N=2,D=4$ supergravity it has
  been shown how the local supersymmetry algebra can be closed on some of
  these dual 2-form fields \cite{Bergshoeff:2007ij}.} In general dimensions
$p$-form potentials of even higher rank are introduced. For instance, the
action corresponding to certain gaugings in $D=6$ requires magnetic 2-form and
3-form potentials \cite{Bergshoeff:2007ef}. This led to the notion of a
\textit{tensor hierarchy}, which consists of a system of potentials of all
degrees $p=1,\ldots, D$ and their respective curvatures, which are related by
Bianchi identities. Note that the tensor hierarchy does not contain $0$-form
potentials, i.e.~scalars, and the metric tensor. These are introduced at a
later stage, see below.

We wish to stress that for theories in specific dimensions generically not the
full tensor hierarchy is used or needed in the construction of an
action. Moreover, the field equations for the new (magnetic) potentials take
the form of \textit{projected} duality relations and, therefore, do not encode
the full set of second-order equations via their integrability conditions. It
is the purpose of this paper to investigate gauged supergravities from the
point of view that all bosonic field equations (except the Einstein equation
and part of the scalar equations of motion) should be derivable from
first-order duality relations. This will naturally include the \textit{full}
tensor hierarchy, which is required by consistency. We will focus on the
bosonic gauge symmetries that are realized by the $D=3$ and (non-anomalous\footnote{By a non-anomalous tensor hierarchy we refer to a specific form of the so-called representation (or linear) constraint imposed on the
embedding tensor. This constraint is such that the classical action of the corresponding gauged
supergravity is gauge invariant.})
$D=4$ tensor hierarchy independent of any supersymmetry. Our results apply
for any number of supersymmetries, not just the maximal or half-maximal
cases. Hence we obtain an off-shell formulation\,\footnote{By ``off-shell
  formulation'' we mean that the commutator algebra of gauge symmetries closes
  without the need to impose constraints on the fields. In this sense an
  off-shell formulation is not related to any particular action.} of all
bosonic symmetries that act in the bosonic sector of any (non-anomalous)
$D=3,4$ gauged supergravity theory.

In the $D=4$ case we use as our staring point Ref.~\cite{deWit:2005ub}. We use
the same formalism, impose the same constraints on the embedding tensor and
follow the same steps up to the 2-form level reproducing exactly the same
results, but we carry out the program to its completion, determining
explicitly all the 3- and 4-forms and their gauge transformations. Here we
find already a surprise in the sense that in $D=4$ we find more top-form
potentials than follow from the expectations formulated in
Refs.~\cite{deWit:2008gc,Bergshoeff:2007vb}\footnote{For instance, we find in
  $D=4$ not only top-forms that correspond to quadratic constraints of the
  embedding tensor but also top-forms that are related to certain linear
  constraints, see subsection \ref{sec-4-form}.}. Our results and the general
results and conjectures of these references\footnote{There are no direct
  computations of tensor hierarchies up to the 4-form level in the
  literature. All we know about them, up to now, is based on general
  arguments.} cannot be straightforwardly compared, though, since in these
works on the general structure of tensor hierarchies only one possible
constraint on the embedding tensor (the \textit{standard} quadratic
constraint) is considered, while in the 4-dimensional setup of
Ref.~\cite{deWit:2005ub} the embedding tensor is subject to two additional
constraints, one quadratic and one linear. They are ultimately responsible for
the existence of additional 4-forms, which we find to be in one-to-one
correspondence with the constraints\footnote{Note added in proof: it has
  recently been shown in Ref.~\cite{Hartong:2009az} that the introduction of
  these additional 4-forms is consistent with $N=1,D=4$
  supergravity. Furthermore, it has been shown that the gauging of particular
  classes of theories (e.g.~$N=1,D=4$ supergravity with a non-vanishing
  superpotential) may require additional constraints on the embedding tensor,
  which lead to extensions of the tensor hierarchy and, in particular, to
  additional 4-forms related to the new constraints.}.

Next, we will make precise how a set of dynamical equations can be defined by
the introduction of first-order duality relations. Besides the $p$-form
potentials these duality relations also contain the scalars and the metric
tensor defining the theory. The set of dynamical equations not only contains
the equations of motion putting all electric potentials on-shell but it also
involves the (projected) scalar equations of motion.  The tensor hierarchy
supplemented by this set of duality relations will be called the
\textit{duality hierarchy}. This set of duality relations cannot be derived
from an action, though the relation to a possible action will be elucidated in
a last step.

For the readers' convenience we briefly outline our program, which
can be summarized by the following 3-step procedure. The first step
consists of the general construction of the tensor hierarchy, which
is an \textit{off-shell} system. The structure in generic dimension
has been given in \cite{deWit:2008ta,deWit:2008gc}. The explicit
form, however, of the {\it
  complete} $D=4$ tensor hierarchy is not available in the literature since it
  was constructed in \cite{deWit:2005ub} only up to the 2-form level. (For the
  construction of the tensor hierarchy of maximal and half-maximal
  4-dimensional supergravities, see \cite{Weidner:2006rp} and references
  therein.)
The complete $D=3$ tensor hierarchy has been discussed in
\cite{deWit:2008ta,Bergshoeff:2008qd}. To construct the tensor
hierarchy one usually starts from the $p$-form potential fields of
all degrees $p=1,\ldots, D$ and then constructs the gauge-covariant
field strengths of all degrees $p=2,\ldots, D$. These field
strengths are related to each other via a set of Bianchi identities
of all degrees $p=3, \ldots,D$. Usually, one starts with the
construction of the covariant field strength for 1-form potentials
which, for general gaugings, requires the introduction of 2-form
potentials. The corresponding 3-form Bianchi identity relates the
2-form field strength to a 3-form field strength for the 2-form
potential, whose construction requires the introduction of a 3-form
potential, etc.  This bootstrap procedure ends with the introduction
of the top-form potentials. The only input required for this
construction is the number of electric $p \ge 1$-form potentials,
the global symmetries of the theory and the representations of this
group under which the $p$-forms transform. Changing these data leads
to different theories that can be seen as different
\textit{realizations} of the low-rank sector of the same tensor
hierarchy.

A trick that simplifies the construction outlined above and which makes the
construction of the complete $D=4$ tensor hierarchy feasible is to first
construct the set of all Bianchi identities relating the $(p+1)$-form field
strengths to the $(p+2)$-field strengths. This systematic construction of the
Bianchi identities can be carried out \textit{even if we do not know
  explicitly the transformation rules of the potentials}. These can be found
afterwards by using the covariance of the different field strengths.  The
resulting gauge transformations form an algebra that closes off-shell: at no
stage of the calculation equations of motions are involved.

The second step is to complement the tensor hierarchy with a set of duality
relations and as such to promote it to what we have called duality
hierarchy.  The duality relations contain more `external' information about
the particular theory we are dealing with. It will introduce the scalars and
the metric tensor field that were not involved in the construction of the
tensor hierarchy\footnote{The dual scalars, i.e.  the $(D-2)$-form potentials,
  are included in the tensor hierarchy.}.  More precisely, some of the duality
relations contain the scalar fields via functions that define all scalar
couplings, i.e.~the Noether currents, the (scalar derivative of the) scalar
potential and functions that define the scalar-vector couplings. In this way
the duality hierarchy contains all the information about the particular
realization of the tensor hierarchy as a field theory.

The duality hierarchy leads to a set of dynamical equations that not only
contains the equations of motion for the electric potentials but it also
involves the (projected) scalar equations of motion according to the rule:
\begin{eqnarray}
\label{eq:rule2} \hskip -.7truecm \mathrm{Tensor\,\, hierarchy}
\;\;\&\;\;\mathrm{duality\,\, relations} &\Leftrightarrow&
\mathrm{dynamical\,\, equations}\, .
\end{eqnarray}
The gauge algebra of the tensor hierarchy closes off-shell even in the
presence of the duality relations. However, in the context of the duality
hierarchy this is a basis-dependent statement. We are free to modify the gauge
transformations by adding terms that are proportional to the duality
relations. Of course, in this new basis the gauge algebra will close on-shell,
i.e.~up to terms that are proportional to the duality relations. We will call
the original basis with off-shell closed algebra the off-shell basis.

The last and third step is the construction of a gauge-invariant
action for all $p$-form potentials, scalars and
metric.\footnote{Strictly speaking, in $D=4$ not all 2-forms enter
the action, see sec.~\ref{sec-4daction}.} In this last step we
encounter a few subtleties that we will clarify. In particular, we
will answer the following questions:
\begin{enumerate}
\item How are the equations of motion that follow from the gauge-invariant
  action related to the set of dynamical equations defined by the duality
  hierarchy?
\item How are the gauge transformations of the $p$-form potentials occurring
  in the action related to the gauge transformations that follow from the
  tensor hierarchy?
\end{enumerate}
It turns out that the construction of a gauge-invariant action requires that
the gauge transformations of the duality hierarchy are given in a particular
basis that can be obtained from the off-shell basis by a change of basis that
will be described in this paper. To be specific, the two sets of transformation rules
(those corresponding to the off-shell tensor hierarchy and those that leave the action invariant)
differ by terms that are proportional to the duality relations.
It is important to note that once a
gauge-invariant action is specified the gauge transformations that leave this
action invariant are not anymore related to the off-shell basis by a
legitimate basis transformation {\sl from the action point of view}.  This is
due to the fact that from the action point of view one is not allowed to remove terms
that are {\sl not} proportional to one of the equations of motion that follow from this action\,\footnote{One
may only change the gauge transformations by adding so-called ``equations of motion symmetries''.}.
 However, although some projected duality relations follow by
extremizing the action, this is not the case for all duality relations of the
duality hierarchy. Therefore, from the action point of view, the gauge
transformations that leave the action invariant are not equivalent to the
gauge transformations of the duality hierarchy in the off-shell basis.
Indeed, the gauge transformations in the off-shell basis do not leave the action
invariant.

This work is organized as follows. In section~\ref{embedding} we briefly
review a few basic facts about the embedding tensor formalism that will be
needed later on. In section~\ref{sec-d4} we construct the complete $D=4$
tensor hierarchy for non-anomalous supergravities. We introduce the setup of
our procedure in subsection~\ref{sec-setup}, present the standard construction
of the vector 2-form field strengths in subsection~\ref{sec-2-form} and the
construction of the 3-form and 4-form field strengths in
subsections~\ref{sec-3-form} and \ref{sec-4-form}. In
section~\ref{sec-d4dualityhierarchy} we add to the tensor hierarchy duality
relations and construct the duality hierarchy. We show how the set of
dynamical equations that follows from this duality hierarchy not only contains
the equations of motion of the different potentials but also the (projected)
scalar equations of motion. Finally, in section~\ref{sec-4daction} we
construct a gauge-invariant action for all the fields of the $D=4$ tensor
hierarchy and show how this result is related to the duality hierarchy. The
general analysis of theories in $D=3$ is presented in Section~\ref{sec-3d},
which completes the investigations \cite{deWit:2008ta,Bergshoeff:2008qd}
discussed in the literature so far.
The $D=4$ and $D=3$ cases may be studied independently, and the latter serves
as a toy model which elucidates some (but not all) of the subtleties of the
four-dimensional analysis. Our conclusions are contained in
section~\ref{sec-conclusions} and the three appendices contain a summary of
the 4-dimensional results.


\section{The embedding tensor formalism}
\label{embedding}

We start by giving a brief review of the the embedding tensor formalism
\cite{Nicolai:2000sc,Nicolai:2001sv,deWit:2008ta,deWit:2008gc}.  Readers
familiar with this technique may skip this part.


\label{generalities}

The embedding tensor formalism is a convenient tool to study gaugings of
supergravity theories in a universal and general way, that does not require a
case-by-case analysis. This technique formally maintains covariance with
respect to the global invariance group $G$ of the ungauged theory, even though
in general $G$ will ultimately be broken by the gauging to the subgroup that
is gauged.  It turns out that all couplings that deform an ungauged
supergravity into a gauged one, as Yukawa couplings, scalar potentials, etc.,
can be given in terms of a special tensor, called the embedding tensor.  Thus,
gauged supergravities are classified by the embedding tensor, subject to a
number of algebraic or group-theoretical constraints, some of which we will
discuss below.

To be more precise, the embedding tensor $\Theta_{M}{}^{\alpha}$ pairs the
generators $t_{\alpha}$ of the group $G$ with the vector fields
$A_{\mu}{}^{M}$ used for the gauging.  The indices $\alpha,\beta,\ldots$ label
the adjoint representation of $G$ and the indices $M,N,\ldots$ label the
representation $\mathcal{R}_{V}$ of $G$, in which the vector fields that will
be used for the gauging transform. Thus, the choice of
$\Theta_{M}{}^{\alpha}$, which generally will not have maximal rank,
determines which combinations of vectors

\begin{equation}
A_{\mu}{}^{M}\Theta_{M}{}^{\alpha}\, ,
\end{equation}

\noindent
can be seen as the gauge fields associated to (a subset of) the generators
$t_{\alpha}$ of the group $G$, and, simultaneously, or alternatively, which
combinations of group generators

\begin{eqnarray}
X_{M} \ = \ \Theta_{M}{}^{\alpha}\;t_{\alpha}\;
\end{eqnarray}

\noindent
can be seen as the generators of the gauge group. Consequently, the embedding
tensor can be used to define covariant derivatives

\begin{eqnarray}
D_{\mu} \ = \
\partial_{\mu}-A_{\mu}{}^{M}\;\Theta_{M}{}^{\alpha}\;t_{\alpha}
\ = \ \partial_{\mu}-A_{\mu}{}^{M}\;X_{M}\; ,
\end{eqnarray}

\noindent
which shows that the embedding tensor can also be interpreted as a set of
gauge coupling constants\footnote{$G$ may have a product structure and each
  factor may have a different coupling constant, which is contained in the
  embedding tensor. We, therefore, do not write any other explicit coupling
  constants apart from $\Theta_{M}{}^{\alpha}$.} of the theory.  Even though
$\Theta_{M}{}^{\alpha}$ has been introduced as a tensor of the duality group
$G$, it is not taken to transform according to its index structure, i.e.~in
the tensor product $\mathcal{R}_{V}\otimes \text{Adj}^{*}$, but must be inert
under $G$ for consistency. This requirement leads to the so-called quadratic
constraints, which state that the embedding tensor is invariant under the
gauge group.  If we denote the generators of $G$ (with structure constants
$f_{\alpha\beta}{}^{\gamma}$) in the representation $\mathcal{R}_{V}$ by
$(t_{\alpha})_{M}{}^{N}$, this amounts to the condition

\begin{equation}
\label{quadconstr} \delta_{ P}\Theta_{M}{}^{\alpha} \ = \
\Theta_{P}{}^{\beta} t_{\beta
M}{}^{N}\Theta_{N}{}^{\alpha}+\Theta_{P}{}^{\beta}f_{\beta\gamma}{}^{\alpha}
\Theta_{M}{}^{\gamma} \ = \ 0\;.
\end{equation}

\noindent
Therefore, seemingly $G$-covariant expressions actually break the duality
group to the subgroup which is gauged.

In the next sections  we will frequently make use of the objects

\begin{equation}
\label{Xtensor} X_{M N}{}^{P} \ \equiv \
\Theta_{M}{}^{\alpha}t_{\alpha N}{}^{P} \ = \ X_{[ M
N]}{}^{P}+Z^{P}{}_{M N}\;,
\end{equation}

\noindent
with $Z^{P}{}_{M N}$ denoting the symmetric part of $X_{M N}{}^{P}$, in terms
of which the quadratic constraints read

\begin{equation}
\label{quadconstr2} \Theta_{P}{}^{\alpha}Z^{P}{}_{MN} \ = \ 0\;.
\end{equation}

\noindent
Thus, the antisymmetry of the `structure constants' of the gauge group holds
only upon contraction with the embedding tensor. Similar relations, that are
familiar from ordinary gauge theories but hold in the present context only
upon contraction with $\Theta$, will be encountered at several places in the
next sections. Note that standard closure of the gauge group follows from
(\ref{quadconstr}) in that

\begin{eqnarray}\label{standardclosure}
\left[ X_{M},X_{N}\right] \ = \ -X_{M N}{}^{P}X_{P} \ = \
-X_{[MN]}{}^{P}X_{P}
\end{eqnarray}

\noindent
by virtue of (\ref{quadconstr2}).

So far, the discussion has been quite general. In the remaining part of this
paper we are going to discuss the $D=4$ and $D=3$ tensor hierarchies in full
detail. For these cases the embedding tensor can be specialized according to
the known representation of the vector fields. Also, our notation for the
indices will slightly differ from the general case to accord with the
literature. In the $D=4$ case we will work with electric vectors
$A^\Lambda{}_\mu$, with $\Lambda = 1, \ldots,\bar n$, and magnetic vectors
$A_{\Lambda\mu}$. Together, these vectors will be combined into a symplectic
contravariant vector $A^M{}_\mu$ with $M$ labeling the fundamental
representation of $Sp(2\bar n, \mathbb{R})$. Also the adjoint index of the
global symmetry group will be denoted by $A$ instead of $\alpha$. This leads
to the following notation for the $D=4$ embedding tensor:
\begin{equation}
D=4\,:\hskip 1truecm \Theta_{M}{}^\alpha\ \ \  \rightarrow\ \ \
\Theta_M{}^A\,.
\end{equation}
On the other hand, in the $D=3$ case the representation $\mathcal{R}_V$ of the
vector fields is equal to the adjoint representation of the global symmetry
group $G$. Therefore, the $D=3$ embedding tensor carries two adjoint indices
and this leads to the following notation\,:\footnote{We assume that $G$
  carries an invariant Cartan-Killing form, such that the indices can be
  freely raised and lowered. This assumption is satisfied for the duality
  groups of three-dimensional supergravity.}
\begin{equation}
D=3\,:\hskip 1truecm \Theta_{M}{}^\alpha\ \ \  \rightarrow\ \ \
\Theta_{M N}\,.
\end{equation}
We now discus the $D=4$ tensor hierarchy  in sections \ref{sec-d4},
\ref{sec-d4dualityhierarchy} and \ref{sec-4daction} and, next, the
$D=3$ tensor hierarchy in section \ref{sec-3d}.

\section{The $D=4$ tensor hierarchy}
\label{sec-d4}

In this section we will construct the complete $D=4$ tensor
hierarchy extending the results of Ref.~\cite{deWit:2005ub}
following the outline of Ref.~\cite{deWit:2008ta}. We will  follow
closely the notation and conventions used in these references.


\subsection{The setup}
\label{sec-setup}

The (bosonic) electric fields of any 4-dimensional field theory are
the metric, scalars and (electric) vectors. Only the latter are
needed in the construction of the tensor hierarchy. We  denote them
by $A^{\Lambda}{}_{\mu}$ where
$\Lambda,\Sigma,\ldots=1,\cdots,\bar{n}$.  In 4-dimensional ungauged
theories one can always introduce their magnetic duals which we
denote by a similar index in lower position $A_{\Lambda\, \mu}$.

The symmetries of the equations of motion of 4-dimensional theories
that act on the electric and magnetic vectors are always subgroups
of $Sp(2\bar{n},\mathbb{R})$ \cite{Gaillard:1981rj} . Thus, it is
convenient to define the symplectic contravariant vector

\begin{equation}
  A^{M}{}_{\mu} \ = \ \left(
  \begin{array}{c}
  A^{\Lambda}{}_{\mu} \\ A_{\Lambda\, \mu} \\
  \end{array} \right)\;.
\end{equation}

\noindent
It is also convenient to define the symplectic metric $\Omega_{MN}$ by

\begin{equation}
\Omega_{MN}=\left(
\begin{array}{cc}
0 & \mathbb{I}_{\bar{n}\times\bar{n}} \\
-\mathbb{I}_{\bar{n}\times\bar{n}} & 0 \\
\end{array}
\right)\, ,
\end{equation}

\noindent and its inverse $\Omega^{MN}$ by

\begin{equation}
\Omega^{MN}\Omega_{NP}=-\delta^{M}{}_{P}\, .
\end{equation}

\noindent
They will be used, respectively, to lower and raise
symplectic indices, \textit{e.g.}\footnote{In what follows we will
mostly use differential-form
  language and suppress the spacetime indices.}

\begin{equation}
A_{M} \equiv \Omega_{MN}A^{N} = (A_{\Lambda}\, , -A^{\Lambda})\, ,
\hspace{1cm} A^{M} = A_{N}\Omega^{NM}\, .
\end{equation}

\noindent The contraction of contravariant and covariant symplectic
indices is, evidently, equivalent to the symplectic product:
$A^{M}B_{M}=A^{M}\Omega_{MN}B^{N} = -A_{M}B^{M}$.

We denote the global symmetry group of the theory by $G$ and its
generators by $T_{A}$, $A,B,C,\ldots=1,\cdots,\mathrm{rank}\, G$.
These satisfy the commutation relations

\begin{equation}
[T_{A},T_{B}] = -f_{AB}{}^{C}T_{C}\, .
\end{equation}

\noindent
$G$ can actually be larger than $Sp(2\bar{n},\mathbb{R})$
and/or not be contained in it\footnote{The symmetries of a set of
scalars decoupled from the
  vectors are clearly unconstrained.}, but, according to the above discussion,
it will always act on $A^{M}$ as a subgroup of it,
i.e.~infinitesimally

\begin{equation}
\delta_{\alpha}A^{M}=\alpha^{A}T_{A\, N}{}^{M}A^{N}\, , \hspace{1cm}
\delta_{\alpha}A_{M}=-\alpha^{A}T_{A\, M}{}^{N}A_{N}\, ,
\end{equation}

\noindent where

\begin{equation}
\label{eq:Tsymplectic} T_{A\, [MN]}\equiv T_{A\,
[M}{}^{P}\Omega_{N]P} =0\, .
\end{equation}

\noindent
This is an important general property of the 4-dimensional case. It is
implicit in this formalism that some of the matrices $T_{A\, M}{}^{N}$ may act
trivially on the vectors, i.e.~they may vanish. Otherwise we could only deal
with $G \subset Sp(2\bar{n},\mathbb{R})$.

Apart from its global symmetries, an ungauged theory containing
$\bar{n}$ Abelian vector fields will always be invariant under the
$2\bar{n}$ Abelian gauge transformations

\begin{equation}
\delta_{\Lambda}A^{M}{}_{\mu} =-\partial_{\mu}\Lambda^{M}\, ,
\end{equation}

\noindent
where $\Lambda^{M}(x)$ is a symplectic vector of local gauge parameters.

To gauge a subgroup of the global symmetry group $G$ we must promote the
global parameters $\alpha^{A}$ to arbitrary spacetime functions
$\alpha^{A}(x)$ and make the theory invariant under these new
transformations. This is achieved by identifying these arbitrary functions
with a subset of the (Abelian) gauge parameters $\Lambda^{M}$ of the vector
fields and subsequently using the corresponding vectors as gauge fields. This
identification is made through the embedding tensor $\Theta_{M}{}^{A}\equiv
(\Theta_{\Lambda}{}^{A}\, , \Theta^{\Lambda\, A})$:

\begin{equation}
\alpha^{A}(x)\equiv \Lambda^{M}(x)\Theta_{M}{}^{A}\, .
\end{equation}

\noindent
The embedding tensor allows us to keep treating all vector fields, used for
gaugings or not, on the same footing. It hence allows us to formally preserve
the symplectic invariance even after gauging.

As discussed in section \ref{embedding} the embedding tensor must satisfy a
number of constraints which guarantee the consistency of the theory. Some of
these constraints have already been discussed in section \ref{embedding}. In
total we have three constraints which we list below. First of all, in the
$D=4$ case we must impose the following quadratic constraint

\begin{equation}
\label{eq:quadraticE} Q^{AB}\equiv
{\textstyle\frac{1}{4}}\Theta^{M\, [A}  \Theta_{M}{}^{B]} =0\, ,
\end{equation}

\noindent
which guarantees that the electric and magnetic gaugings
are mutually local \cite{deWit:2005ub}. Observe that the
antisymmetry of $\Omega^{MN}$ and the above constraint imply
$\Theta^{M\, A} \Theta_{M}{}^{B}=0$. This constraint is a particular
feature of the 4-dimensional case.

As mentioned in section \ref{embedding} there is a second quadratic
constraint which encodes the fact that the embedding tensor has to
be itself invariant under gauge transformations. If the gauge
transformations of objects with contravariant and covariant
symplectic indices are

\begin{equation}
  \delta_{\Lambda}\xi^{M}=\Lambda^{N} \Theta_{N}{}^{A}T_{A\, P}{}^{M}\xi^{P}\, ,
  \hspace{1cm}
  \delta_{\Lambda} \eta_{M}=-\Lambda^{N} \Theta_{N}{}^{A} T_{A\, M}{}^{P}\xi_{P}\, ,
\end{equation}

\noindent
and the gauge transformations of objects with
contravariant and covariant adjoint indices are written in the form

\begin{equation}
\delta_{\Lambda}\pi^{A}=
\Lambda^{M}\Theta_{M}{}^{B}f_{BC}{}^{A}\pi^{C}\, . \hspace{1cm}
\delta_{\Lambda}\zeta_{A}=
-\Lambda^{M}\Theta_{M}{}^{B}f_{BA}{}^{C}\zeta_{C}\, ,
\end{equation}

\noindent
then

\begin{equation}
\label{eq:quadraticTdef} \ \delta_{\Lambda}\Theta_{M}{}^{A} =
-\Lambda^{N} Q_{NM}{}^{A}\, , \hspace{1cm} Q_{NM}{}^{A} \equiv
\Theta_{N}{}^{A}T_{A\,  M}{}^{P}\Theta_{P}{}^{A}
-\Theta_{N}{}^{A}\Theta_{M}{}^{B}f_{AB}{}^{A}\, ,
\end{equation}

\noindent
and the second quadratic constraint reads

\begin{equation}
\label{eq:quadraticT}
Q_{NM}{}^{A}=0\, .
\end{equation}

The third constraint applies to all 4-dimensional supergravity theories that
are free of gauge anomalies \cite{DeRydt:2008hw} and can be expressed using
the $X$ generators introduced in section \ref{embedding}, see
Eq.~\eqref{Xtensor}:

\begin{equation}
\label{eq:Xdef}
X_{M} \equiv \Theta_{M}{}^{A}T_{A}\, ,
\hspace{1cm}
X_{MN}{}^{P} \equiv \Theta_{M}{}^{A}T_{A\, N}{}^{P}\, .
\end{equation}

\noindent
This constraint (the so-called \textit{representation constraint}) is linear
in $\Theta_{M}{}^{A}$ and reads as follows \cite{deWit:2005ub}:

\begin{equation}
\label{eq:linear} L_{MNP}\equiv X_{(MNP)} =
X_{(MN}{}^{Q}\Omega_{P)Q}=0\, .
\end{equation}

The three constraints that the embedding tensor has to satisfy are not
independent, but are related by

\begin{equation}
\label{eq:reconstraint2}
Q_{(MN)}{}^{A} -3L_{MNP}Z^{PA} -2Q^{AB}T_{BMN}=0\, .
\end{equation}
This relation can be used to show that the constraint $Q^{AB}=0$ follows from
the constraint $Q_{(MN)}{}^{A}=0$ when the linear constraint $L_{MNP}=0$ is
explicitly solved, whenever the action of the global symmetry group on the
vectors is faithful. We will neither solve explicitly the linear constraint by
choosing to work only with representations allowed by it, nor we will assume
the action of the global group on the vectors to be faithful, since there are
many interesting situations in which this is not the case and we aim to
be as general as possible. In (half-) maximal supergravities, though, the
global symmetry group always acts faithfully on the vector fields.

These two choices, which differ from those made in the explicit examples found
in the literature (see e.g.~Ref.~\cite{Weidner:2006rp}) will have important
consequences in the field content of the tensor hierarchy and are the reason
why our results also differ from those obtained in them.

\noindent
Before we go on we wish to collect a few properties of the $X$ generators
$X_{MN}{}^P$ in a separate subsection.


\subsubsection{The $X$ generators and their properties}

We first discuss the symmetry properties of the $X$ generators. By their
definition, and due to the symplectic property of the $T_{A\, N}{}^{P}$
generators, see Eq.~(\ref{eq:Tsymplectic}), we have

\begin{equation}
X_{MNP}=X_{MPN}\, .
\end{equation}

\noindent
From the definition of the quadratic constraint Eq.~(\ref{eq:quadraticT}) it
follows that

\begin{equation}
X_{(MN)}{}^{P}\Theta_{P}{}^{C} = Q_{(MN)}{}^{C}\, ,
\end{equation}

\noindent and so it will vanish\footnote{Here we will keep the terms
proportional to constraints for later use, including the linear
constraints in (\ref{eq:symmproperty}).}, although, in general, we
will have

\begin{equation}
X_{(MN)}{}^{P} \neq 0\, .
\end{equation}

\noindent This implies, in particular

\begin{equation}
\label{eq:symmproperty} X_{(MN)P}= -{\textstyle\frac{1}{2}}X_{PMN}
+{\textstyle\frac{3}{2}}L_{MNP}\,\,\, \Rightarrow X_{(MN)}{}^{P} =
Z^{PA}T_{AMN}+{\textstyle\frac{3}{2}}L_{MN}{}^P\, ,
\end{equation}

\noindent
where we have defined

\begin{equation}
Z^{PA} \equiv -{\textstyle\frac{1}{2}}\Omega^{NP}\Theta_{N}{}^{A} =
\left\{
  \begin{array}{l}
    +{\textstyle\frac{1}{2}}\Theta^{\Lambda A}\, ,\\
\\
    -{\textstyle\frac{1}{2}}\Theta_{\Lambda}{}^{A}\, ,\\
  \end{array}
\right. \, .
\end{equation}

\noindent $Z^{PA}$ will be used to project in directions orthogonal
to the embedding tensor since, due to the first quadratic constraint
Eq.~(\ref{eq:quadraticE}), we find that

\begin{equation}
Z^{MA}\Theta_{M}{}^{B}=-{\textstyle\frac{1}{2}}Q^{AB}\, .
\end{equation}

We next discuss some properties of the products of two $X$ generators.  From
the commutator of the $T_{A}$ generators and the definition of the generators
$X_{M}$ and the matrices $X_{MN}{}^{P}$ we find the commutator of the $X_{M}$
generators to be

\begin{equation}
\label{eq:unconstrainedXcommutator} [X_{M},X_{N}] =
Q_{MN}{}^{C}T_{C}-X_{MN}{}^{P}X_{P}\, .
\end{equation}

\noindent
This reduces to (cf. to Eq. \eqref{standardclosure})

\begin{equation}
\label{eq:constrainedXcommutator} [X_{M},X_{N}] =
-X_{[MN]}{}^{P}X_{P}\, ,
\end{equation}

\noindent
upon use of the above constraint and $Q_{MN}{}^{C}=0$.  From the commutator
Eq.~(\ref{eq:unconstrainedXcommutator}) one can derive the analogue of the
Jacobi identities

\begin{equation}
\label{eq:Jacobi}
\begin{array}{rcl}
X_{[MN]}{}^{Q}X_{[PQ]}{}^{R} + X_{[NP]}{}^{Q}X_{[MQ]}{}^{R}
+X_{[PM]}{}^{Q}X_{[NQ]}{}^{R} & = &
\\
& & \\
& & \hspace{-7cm} =-{\textstyle\frac{1}{3}}
\{X_{[MN]}{}^{Q}X_{(PQ)}{}^{R} + X_{[NP]}{}^{Q}X_{(MQ)}{}^{R}
+X_{[PM]}{}^{Q}X_{(NQ)}{}^{R}\}
\\
& & \\
& & \hspace{-6.5cm} -Q_{[MN|}{}^{C}T_{C\, |P]}{}^{R}\, .
\end{array}
\end{equation}

We finally present two more useful identities that can be derived from the
commutators:

\begin{eqnarray}
X_{(MN)}{}^{Q}X_{PQ}{}^{R} -X_{PN}{}^{Q}X_{(MQ)}{}^{R}
-X_{PM}{}^{Q}X_{(NQ)}{}^{R} & = & -Q_{P(M|}{}^{C}T_{C\, |N)}{}^{R}\, ,\\
& & \nonumber \\
X_{[MN]}{}^{Q}X_{PQ}{}^{R} -X_{PN}{}^{Q}X_{[MQ]}{}^{R}
+X_{PM}{}^{Q}X_{[NQ]}{}^{R} & = & Q_{P[M|}{}^{C}T_{C\, |N]}{}^{R}\,
.
\end{eqnarray}


\subsection{The vector field strengths $F^{M}$}
\label{sec-2-form}

We now return to the construction of the field strengths of the different
$p$-form potentials. In what follows we will set all the constraints
explicitly to zero in order to simplify the expressions. In this section we
consider the vector field strengths.

To construct the vector field strength it is convenient to start from the
covariant derivative. This derivative acting on objects transforming according
to $\delta\phi=\Lambda^{M}\delta_{M}\phi$ is defined by

\begin{equation}
\mathfrak{D}\phi = d\phi +A^{M}\delta_{M}\phi\, .
\end{equation}

\noindent
For instance, the covariant derivative of a contravariant symplectic vector

\begin{equation}
\mathfrak{D}\xi^{M} =d\xi^{M}  +X_{NP}{}^{M}A^{N}\xi^{P}\, ,
\end{equation}

\noindent
transforms covariantly provided that

\begin{equation}
\label{eq:Astandardgaugetrans} \delta A^{M} =
-\mathfrak{D}\Lambda^{M} +\Delta A^{M}\, , \hspace{1cm}
\Theta_{M}{}^{A}\Delta A^{M}=0\, .
\end{equation}

The Ricci identity of the covariant derivative on $\Lambda^{N}$ can be written
in the form

\begin{eqnarray}
\mathfrak{D}  \mathfrak{D}  \Lambda^{M} & = &
X_{NP}{}^{M}F^{N}\Lambda^{P}\, ,
\end{eqnarray}

\noindent
for some 2-form $F^{M}$. Since this expression is gauge-covariant, $F^{M}$, contracted with the embedding tensor,
will automatically be gauge-covariant, whatever it is and it is natural to
identify it with the gauge-covariant vector field strength. The above
expression defines it up to a piece $\Delta F^{M}$ which is projected out by
the embedding tensor, just like $\Delta A^{M}$ in $\delta A^M$. An explicit
calculation gives

\begin{equation}
F^{M} = dA^{M} +{\textstyle\frac{1}{2}}X_{[NP]}{}^{M}A^{N}\wedge
A^{P} +\Delta F^{M}\, , \hspace{1cm} \Theta_{M}{}^{A}\Delta
F^{M}=0\, .
\end{equation}

The possible presence of $\Delta F^{M}$ is a novel feature of the embedding
tensor formalism. Its gauge transformation rule can be found by using the
gauge covariance of $F^{M}$. Under Eq.~(\ref{eq:Astandardgaugetrans}), using
$\Theta_{M}{}^{A}\Delta F^{M}=0$, we find that

\begin{equation}
\delta F^{M}
=
\Lambda^{P}X_{PN}{}^{M}F^{N} +\mathfrak{D}\Delta A^{M}
-2 X_{(NP)}{}^{M}(\Lambda^{N}F^{P}
+{\textstyle\frac{1}{2}}A^{N}\wedge \delta A^{P}) +\delta\Delta
F^{M}\, ,
\end{equation}

\noindent
so that $F^M$ transforms covariantly provided that we take

\begin{equation}
\delta\Delta F^{M}
=
-\mathfrak{D}\Delta A^{M} +2 Z^{MA} T_{A\,
NP}(\Lambda^{N}F^{P} +{\textstyle\frac{1}{2}}A^{N}\wedge \delta
A^{P})\, ,
\end{equation}

\noindent
where we have used Eq.~(\ref{eq:symmproperty}). Since both $\Delta A^{M}$ and
$\Delta F^{M}$ are annihilated by the embedding tensor, we conclude that in
the generic situation we are considering here\footnote{The only information we
  have about the embedding tensor in a generic situation is provided by the
  three constraints $Q_{NP}{}^{E}=0\, ,\,\,Q^{AB}=0\, ,\,\, L_{MNP}=0$. There
  is only one which we can write in the form $\Theta_{M}{}^{A}\times
  \mathrm{Something}^{M}=0$, which is the constraint $Q^{AB}=0$ and that
  uniquely identifies $\mathrm{Something}^{M}=Z^{MB}$ up to a proportionality
  constant.} $\Delta F^{M} = Z^{MA}B_{A}$ where $B_{A}$ is some 2-form field
in the adjoint of $G$ and $\Delta A^{M} = -Z^{MA}\Lambda_{A}$ where
$\Lambda_{A}$ is a 1-form gauge parameter in the same representation. Then

\begin{eqnarray}
\label{eq:FM} F^{M} & = & dA^{M}
+{\textstyle\frac{1}{2}}X_{[NP]}{}^{M}A^{N}\wedge A^{P}
+Z^{MA}B_{A}\, ,
\\
& & \nonumber \\
\delta A^{M} & = & -\mathfrak{D}\Lambda^{M} -Z^{MA}\Lambda_{A}\, ,
\\
& & \nonumber \\
\delta B_{A} & = &  \mathfrak{D}\Lambda_{A} +2T_{A\,
NP}[\Lambda^{N}F^{P} +{\textstyle\frac{1}{2}}A^{N}\wedge\delta
A^{P}] +\Delta B_{A}\, ,
\end{eqnarray}

\noindent
where $\Delta B_{A}$ is a possible additional term which is projected out by
$Z^{MA}$, i.e.~$Z^{MA}\Delta B_A=0$, and can be determined by studying the
construction of a gauge-covariant field strength $H_{A}$ for the 2-form
$B_{A}$.


\subsection{The 3-form field strengths $H_{A}$}
\label{sec-3-form}

We continue to determine the form of $H_{A}$ using the Bianchi identity for
$F^{M}$ just as we used the Ricci identity to find an expression for $F^{M}$.
An explicit computation using Eq.~(\ref{eq:FM}) gives

\begin{equation}
\mathfrak{D}F^{M} = Z^{MA}\{\mathfrak{D}B_{A} +T_{A\,
RS}A^{R}\wedge[dA^{S}
+{\textstyle\frac{1}{3}}X_{NP}{}^{S}A^{N}\wedge A^{P}]\}\, .
\end{equation}

\noindent
It is clear that the expression in brackets must be covariant and it defines a
3-form field strength $H_{A}$ up to terms $\Delta H_{A}$ that are projected
out by $Z^{MA}$, i.e.

\begin{eqnarray}
\label{eq:bianchiFM}
\mathfrak{D}F^{M} & = & Z^{MA}H_{A}\, ,\\
& & \nonumber \\
H_{A} & = & \mathfrak{D}B_{A} +T_{A\, RS}A^{R}\wedge[dA^{S}
+{\textstyle\frac{1}{3}}X_{NP}{}^{S}A^{N}\wedge A^{P}] +\Delta
H_{A}\,
\end{eqnarray}

\noindent
with $Z^{MA}\Delta H_{A} =0$. Both $\Delta B_{A}$ and $\Delta H_{A}$ are
determined by requiring gauge covariance of $H_{A}$. An explicit calculation
gives

\begin{equation}
\label{eq:esa}
  \begin{array}{rcl}
\delta H_{A} & = & -\Lambda^{M}\Theta_{M}{}^{B}f_{BA}{}^{C} H_{C}
\\
& & \\
& & -Y_{AM}{}^{C}[\Lambda^{M}H_{C} -\delta A^{M}\wedge B_{C}
-F^{M}\wedge \Lambda_{C} -{\textstyle\frac{1}{3}}T_{C\, NP}
A^{M}\wedge A^{N} \wedge \delta A^{P}]
\\
& & \\
& &
+\mathfrak{D}\Delta B_{A}+\delta \Delta H_{A}\, .\\
\end{array}
\end{equation}

\noindent
We have defined the $Y$-tensor as

\begin{equation}
\label{eq:Ytensordef} Y_{AM}{}^{C} \equiv
\Theta_{M}{}^{B}f_{AB}{}^{C} -T_{A\, M}{}^{N}\Theta_{N}{}^{C}\, .
\end{equation}

\noindent
and it satisfies the condition

\begin{equation}
Z^{MA} Y_{AN}{}^{C}= {\textstyle\frac{1}{2}}\Omega^{PM}
Q_{PN}{}^{C}= 0\, .
\end{equation}

The 3-form field strengths $H_{A}$ transform covariantly provided that the
last two lines in Eq.~(\ref{eq:esa}) vanish. A natural solution is to take

\begin{eqnarray}
\Delta B_{A} \equiv -Y_{AM}{}^{C}\Lambda_{C}{}^{M}\, , \hspace{1cm}
\Delta H_{A} \equiv Y_{AM}{}^{C}C_{C}{}^{M}\, ,
\end{eqnarray}

\noindent
where $\Lambda_{C}{}^{M}$ is a 2-form gauge parameter and $C_{C}{}^{M}$ is a
3-form field
about which we will not make any assumptions for the moment. In particular, we
will not assume it to satisfy any constraints in spite of the fact that we
expect it to be ``dual'' to the embedding tensor, which is a constrained
object.  We are going to see that, actually, we are not going to need any such
explicit constraints to construct a fully consistent tensor hierarchy. On the
other hand, we are going to find St\"uckelberg shift symmetries acting on
$C_{C}{}^{M}$ whose role is, precisely, to compensate for the lack of explicit
constraints and, potentially, allow us to remove the same components of
$C_{C}{}^{M}$ which would be eliminated by imposing those constraints. We
anticipate that those St\"uckelberg shift symmetries require the existence of
4-forms in order to construct gauge-covariant 4-form field strengths
$G_{C}{}^{M}$. It should come as no surprise after this discussion, that the
4-forms are in one-to-one correspondence with the constraints of the embedding tensor.
Working with unconstrained fields is simpler and it is one of the advantages
of our approach.

We then, find

\begin{eqnarray}
H_{A} & = & \mathfrak{D}B_{A} +T_{A\, RS}A^{R}\wedge[dA^{S}
+{\textstyle\frac{1}{3}}X_{NP}{}^{S}A^{N}\wedge A^{P}]
+Y_{AM}{}^{C}C_{C}{}^{M}\, ,
\\
& & \nonumber \\
\delta B_{A} & = &  \mathfrak{D}\Lambda_{A} +2T_{A\, NP}
[\Lambda^{N}F^{P}+{\textstyle\frac{1}{2}}A^{N}\wedge \delta A^{P}]
-Y_{AM}{}^{C}\Lambda_{C}{}^{M}\, ,
\\
& & \nonumber \\
\delta C_{C}{}^{M} & = & \mathfrak{D}\Lambda_{C}{}^{M}+
\Lambda^{M}H_{C} -\delta A^{M}\wedge B_{C} -F^{M}\wedge \Lambda_{C}
\nonumber \\
& & \nonumber \\
& & -{\textstyle\frac{1}{3}}T_{C\, NP} A^{M}\wedge A^{N} \wedge
\delta A^{P} +\Delta C_{C}{}^{M}\, ,
\end{eqnarray}

\noindent
where we have introduced a possible additional term
$\Delta C_{C}{}^{M}$ analogous to $\Delta A^{M}$ and $\Delta B_{A}$
which now is projected out by $Y_{AM}{}^{C}$

\begin{equation}
Y_{AM}{}^{C}\Delta C_{C}{}^{M} = 0\, ,
\end{equation}

\noindent
and which will be determined by requiring gauge covariance of the 4-form field
strength $G_{C}{}^{M}$.


\subsection{The 4-form field strengths $G_{C}{}^{M}$}
\label{sec-4-form}

To determine the 4-form field strengths $G_{C}{}^{M}$ we use the Bianchi
identity of $H_{A}$. We can start by taking the covariant derivative of both
sides of the Bianchi identity of $F^{M}$ Eq.~(\ref{eq:bianchiFM}) and then
using the Ricci identity. We thus get

\begin{equation}
Z^{MA}\mathfrak{D}H_{A} = X_{NP}{}^{M}F^{N}\wedge F^{P}=
Z^{MA}T_{A\, NP}F^{N}\wedge F^{P}\, .
\end{equation}

\noindent
This implies that $\mathfrak{D}H_{A} = T_{A\, MN}F^{M}\wedge F^{N}
+\Delta\mathfrak{D}H_{A}$ where $Z^{MA}\Delta\mathfrak{D}H_{A}=0$, suggesting
that $\Delta\mathfrak{D}H_{A} \sim Y_{AM}{}^{C}G_{C}{}^{M}$.  A direct
calculation yields the result

\begin{equation}
  \begin{array}{rcl}
G_{C}{}^{M} & = & \mathfrak{D}C_{C}{}^{M} +F^{M}\wedge B_{C}
-{\textstyle\frac{1}{2}}Z^{MA}B_{A}\wedge B_{C}
\\
& & \\
& & +{\textstyle\frac{1}{3}}T_{C\, SQ} A^{M}\wedge A^{S} \wedge
(F^{Q}-Z^{QA}B_{A})
\\
& & \\
& & -{\textstyle\frac{1}{12}}T_{C\, SQ}X_{NT}{}^{Q}A^{M}\wedge A^{S}
\wedge A^{N} \wedge A^{T}
\\
& & \\
& & +\Delta G_{C}{}^{M}\, ,
\end{array}
\end{equation}

\noindent
where

\begin{equation}
Y_{AM}{}^{C}\Delta G_{C}{}^{M} =0\, .
\end{equation}

\noindent
The Bianchi identity then takes the form

\begin{equation}
\label{eq:bianchiHA} \mathfrak{D}H_{A} = Y_{AM}{}^{C} G_{C}{}^{M}
+T_{A\, MN}F^{M}\wedge F^{N}\, .
\end{equation}

$\Delta C_{C}{}^{M}$ and $\Delta G_{C}{}^{M}$ must now be determined
by using the gauge covariance of the full field strength
$G_{C}{}^{M}$. It is tempting to repeat what we did in the previous
cases. However, the calculation is, now, much more complicated and
it would be convenient to have some information about the new
tensor(s) orthogonal to $Y_{AM}{}^{C}$ that we may expect.

Given that the projectors arise naturally in the computation of the Bianchi
identities, we are going to ``compute'' the Bianchi identity of $G_{C}{}^{M}$
obviating the fact that it is already a 4-form, and in $D=4$ its Bianchi
identity is trivial. We have not used the dimensionality of the problem so far
(except in the existence of magnetic vector fields that gives rise to the
symplectic structure and in the assignment of adjoint indices to the 2-forms)
and, in any case, our only goal in performing this computation is to find the
relevant invariant tensor(s).

Thus, we apply $\mathfrak{D}$ to both sides of Eq.~(\ref{eq:bianchiHA}) using
the Bianchi identity of $F^{M}$ Eq.~(\ref{eq:bianchiFM}) and the Ricci
identity. This leads to the following identity

\begin{equation}
Y_{AM}{}^{C} \{\mathfrak{D}G_{C}{}^{M}  -F^{M}\wedge H_{C}\}=0\, ,
\end{equation}

\noindent
from which it follows that

\begin{equation}
\mathfrak{D}G_{C}{}^{M}  = F^{M}\wedge H_{C} +\Delta
\mathfrak{D}G_{C}{}^{M}\, , \hspace{1cm} Y_{AM}{}^{C}\Delta
\mathfrak{D}G_{C}{}^{M}=0\, .
\end{equation}

Acting again with $\mathfrak{D}$ on both sides of this last equation and using
the Ricci and Bianchi identities, we get in an straightforward manner

\begin{equation}
  \begin{array}{rcl}
\mathfrak{D}\Delta \mathfrak{D}G_{C}{}^{M} & = & W_{C}{}^{MAB}
H_{A}\wedge H_{B}
\\
& & \\
& & +W_{CNPQ}{}^{M} F^{N}\wedge F^{P} \wedge F^{Q}
\\
& & \\
& & +W_{CNP}{}^{EM} F^{N}\wedge G_{E}{}^{P}\, ,
  \end{array}
\end{equation}

\noindent
where

\begin{eqnarray}
\label{eq:W1} W_{C}{}^{MAB} & \equiv & -Z^{M[A}\delta_{C}{}^{B]}\, ,
\\
& & \nonumber \\
\label{eq:W2} W_{CNPQ}{}^{M} & \equiv & T_{C\,
(NP}\delta_{Q)}{}^{M}\, ,
\\
& & \nonumber \\
\label{eq:W3} W_{CNP}{}^{EM} & \equiv &
\Theta_{N}{}^{D}f_{CD}{}^{E}\delta_{P}{}^{M}
+X_{NP}{}^{M}\delta_{C}{}^{E} -Y_{CP}{}^{E}\delta_{N}{}^{M}\, .
\end{eqnarray}

\noindent
We thus found the desired new tensors. The $Y$-tensor annihilates the three
new $W$ tensors in virtue of the 3 constraints satisfied by the embedding tensor

\begin{equation}
Y_{AM}{}^{C}W_{C}{}^{MAB} = Y_{AM}{}^{C}W_{CNPQ}{}^{M}
=Y_{AM}{}^{C}W_{CNP}{}^{EM} =0\, ,
\end{equation}

\noindent
as expected. Note that the first and third $W$-tensors are linear in $\Theta$
but that the second $W$-tensor is independent of $\Theta$.  Other important
sets of identities satisfied by these $W$-tensors can be found in
Appendix~\ref{app-wtensorproperties}.

Coming back to our original problem of determining the form of $\Delta
G_{C}{}^{M}$ and $\Delta C_{C}{}^{M}$, we conclude from the previous analysis
that

\begin{eqnarray}
\Delta C_{C}{}^{M} & = & -W_{C}{}^{MAB}\Lambda_{AB}
-W_{CNPQ}{}^{M}\Lambda^{NPQ} -W_{CNP}{}^{EM}\Lambda_{E}{}^{NP}\,  ,
\\
& & \nonumber \\
\Delta G_{C}{}^{M} & = & W_{C}{}^{MAB}D_{AB} +W_{CNPQ}{}^{M}D^{NPQ}
+W_{CNP}{}^{EM} D_{E}{}^{NP}\,  ,
\end{eqnarray}

\noindent
where $\Lambda_{AB},\Lambda^{NPQ},\Lambda_{E}{}^{NP}$ are 3-form gauge
parameters and $D_{AB},D^{NPQ},D_{E}{}^{NP}$ are possible 4-forms whose
presence will be justified in $G_{C}{}^{M}$ if their gauge transformations are
non-trivial in order to make the 4-form field strengths gauge
covariant.
Taking into account the symmetries of the $W$-tensors, it is easy to see that
$D_{AB}=D_{[AB]}$, $D^{NPQ}=D^{(NPQ)}$ and analogously for the gauge
parameters $\Lambda_{AB},\Lambda^{NPQ}$. $D_{E}{}^{NP}$ and
$\Lambda_{E}{}^{NP}$ have no symmetries.

We observe that the three 4-form $D$-potentials seem to be
associated to the three constraints $Q^{AB}$, $L_{NPQ}$,
$Q_{NP}{}^{E}$ given in Eqs.~(\ref{eq:quadraticE}),
(\ref{eq:quadraticT}) and (\ref{eq:linear}) in the sense that they
carry the same representations. Only the last one was expected
according to the general formalism developed in
Ref.~\cite{deWit:2008ta} and the specific study of the top forms
performed in Ref.~\cite{deWit:2008gc,Bergshoeff:2007vb}. We find
that in 4 dimensions there are more top-form potentials due to the
additional structures (e.g.~the symplectic one) and properties of
4-dimensional theories.

Knowing the different $W$ tensors it is now a relatively
straightforward task to obtain by a direct calculation the expression for
$\delta G_{C}{}^{M}$, collect the terms proportional to the three
$W$-structures and determine the gauge transformations of the three different
4-form $D$-potentials by requiring gauge-covariance of $G_{C}{}^{M}$. An
explicit calculation gives

\begin{eqnarray}
\delta D_{AB} & = & \mathfrak{D}\Lambda_{AB} +\alpha B_{[A}\wedge
Y_{B]P}{}^{E}\Lambda_{E}{}^{P} + \mathfrak{D}\Lambda_{[A}\wedge
B_{B]} -2\Lambda_{[A}\wedge H_{B]}
\nonumber \\
& & \nonumber \\
& &
 +2T_{[A| NP}[\Lambda^{N}F^{P}-{\textstyle\frac{1}{2}}A^{N}\wedge \delta A^{P}]\wedge
B_{|B]}\, , \\
& & \nonumber \\
\delta D_{E}{}^{NP} & = & \mathfrak{D}\Lambda_{E}{}^{NP}
-[F^{N}-{\textstyle\frac{1}{2}}(1-\alpha)Z^{NA}B_{A}]\wedge
\Lambda_{E}{}^{P} +C_{E}{}^{P}\wedge \delta A^{N}
\nonumber \\
& & \nonumber \\
& & +{\textstyle\frac{1}{12}} T_{EQR} A^{N}\wedge A^{P} \wedge A^{Q}
\wedge \delta A^{R}
+\Lambda^{N}G_{E}{}^{P}\, ,\label{deltaDENP}\\
& & \nonumber \\
\delta D^{NPQ} & = & \mathfrak{D}\Lambda^{NPQ} -2A^{(N}\wedge
(F^{P}-Z^{PA}B_{A})\wedge \delta A^{Q)}
 \nonumber \\
& & \nonumber \\
& & +{\textstyle\frac{1}{4}} X_{RS}{}^{(N} A^{P|} \wedge A^{R}\wedge
A^{S} \wedge  \delta A^{|Q)} -3\Lambda^{(N}F^{P}\wedge F^{Q)}\, ,
\end{eqnarray}

\noindent
where $\alpha$ is an arbitrary real constant. We hence find that there is a
1-parameter family of solutions to the problem of finding a gauge-covariant
field strength for the 3-form. The origin of this freedom resides in the
presence of a St\"uckelberg-type symmetry which we discuss in the next
subsection.


\subsubsection{St\"uckelberg symmetries}
\label{sec-stuckelberg}

Differentiating (\ref{eq:reconstraint2}) with respect to $\Theta_{Q}{}^{C}$
using Eqs.~(\ref{eq:dzW1})-(\ref{eq:dzW3}) gives the following identity among
the $W$ tensors:

\begin{equation}
\label{eq:reconstraint3}
W_{C(MN)}{}^{AQ} -3W_{CMNP}{}^{Q}Z^{PA}  -2W_{C}{}^{QAB}T_{B\, MN} =
{\textstyle\frac{3}{2}}
L_{MN}{}^{Q}\delta_{C}{}^{A}\, .
\end{equation}

\noindent The relation (\ref{eq:reconstraint3}) gives rise to
symmetries under St\"uckelberg shifts of the 4-forms in the 4-form
field strength $G_{C}{}^{M}$

\begin{equation}
\label{eq:stuckelberg1}
\begin{array}{rcl}
  \delta D_{E}{}^{NP} & = & \Xi_{E}{}^{(NP)}\, ,\\
  & & \\
  \delta D_{AB} & = & -2 \Xi_{[A}{}^{MN}T_{B]MN}\, ,\\
  & & \\
  \delta D^{NPQ} & = & -3 Z^{(N|A}\Xi_{A}{}^{|PQ)}\, .\\
\end{array}
\end{equation}

\noindent
This shift symmetry, which allows us to remove the part symmetric in $NP$ of
$D_{E}{}^{NP}$, also leaves the 4-form field strengths $G_{C}{}^{M}$
invariant.

If we multiply (\ref{eq:reconstraint2}) by $Z^{NE}$ we find another relation
between constraints

\begin{equation}
\label{eq:reconstraint1}
Q^{AB}Y_{BP}{}^{E}-{\textstyle\frac{1}{2}}Z^{NA}Q_{NP}{}^{E} = 0\, .
\end{equation}

\noindent Differentiating it again with respect to the embedding
tensor we find the following relation between
$W$-tensors\footnote{This identity can also be
  obtained multiplying Eq.~(\ref{eq:reconstraint3}) by $Z^{NE}$.}:

\begin{equation}
\label{eq:reconstraint4}
W_{C}{}^{MAB}Y_{BP}{}^{E}-{\textstyle\frac{1}{2}}Z^{NA}W_{CNP}{}^{EM}=
{\textstyle\frac{1}{4}}Q^{M}{}_{P}{}^{E} \delta_{C}{}^{A}
-Q^{AB}[\delta_{P}{}^{M}f_{BC}{}^{E}-T_{B\, P}{}^{M}\delta_{C}{}^{E}]\, ,
\end{equation}

\noindent which implies that the St\"uckelberg shift

\begin{equation}
\label{eq:stuckelberg2}
\begin{array}{rcl}
\delta D_{E}{}^{NP} & = & {\textstyle\frac{1}{2}}Z^{NB}\Xi_{BE}{}^{P}\, ,\\
& & \\
\delta D_{AB} & = & Y_{[A|P}{}^{E}\Xi_{B]E}{}^{P}\, ,\\
\end{array}
\end{equation}

\noindent leaves invariant the 4-form field strength $G_{C}{}^{M}$
up to terms proportional to the quadratic constraints, which are
taken to vanish identically in the tensor hierarchy.  This shift
symmetry is associated to the arbitrary parameter $\alpha$ in the
gauge transformations of $D_{AB}$ and $D_{E}{}^{NP}$. Observe that,
even though it is based on the identity Eq.~(\ref{eq:reconstraint4})
which we can get from Eq.~(\ref{eq:reconstraint3}), this symmetry is
genuinely independent from that in Eq.~(\ref{eq:stuckelberg1}).

This finishes the construction of the 4-dimensional tensor hierarchy.  The
field strengths, Bianchi identities and gauge transformations of the
hierarchy's $p$-form fields are collected in
Appendix~\ref{app-gaugetranshierarchy}. By construction the algebra of all
bosonic gauge transformations closes off-shell on all $p$-form potentials. No
equations of motion are needed at this stage.


\section{The $D=4$ duality  hierarchy}
\label{sec-d4dualityhierarchy}


In this section we are going to introduce dynamical equations for the tensor
hierarchy via the introduction of first-order duality relations, see
Eq.~(\ref{eq:rule2}). This promotes the tensor hierarchy to a \textit{duality
  hierarchy}. We will see that the dynamical equations will not only contain
the equations of motions of the $p$-form potentials but also the (projected)
scalar equations of motion.  These scalars, together with the metric, will be
introduced via the duality relations.  In particular, the scalar couplings
enter into the duality relations via functions that can be identified with the
Noether currents, the (scalar derivative of the) scalar potential and the
kinetic matrix describing the coupling of the scalars to the vectors. In this
way the duality hierarchy puts the tensor hierarchy on-shell and establishes a
link with a Yang-Mills-type gauge field theory containing a metric, scalars
and $p$-form potentials. This field theory can be viewed as the bosonic part
of a gauged supergravity theory. We stress that at this point we only compare
equations of motion. It is only in the last and third step that we consider an
action for the fields of the hierarchy. We will assume that the
Yang-Mills-type gauge field theory has an action but we will only consider its
equations of motion in order to properly identify in the duality relations the
Noether current, scalar potential and the scalar-vector kinetic function.

In the next subsection we will first consider a Yang-Mills-type
gauge field theory with purely electric gaugings, i.e.~only electric
1-forms are involved in the gauging. In particular we will compare
the equations of motion of this field theory with the dynamical
equations of the duality hierarchy. This example shows us how to
introduce the metric and scalars in the duality hierarchy. In the
next subsection we will first consider a formally
symplectic-covariant generalization of the equations of motion with
purely electric gaugings. This generalization necessarily involves
electric and magnetic gaugings. We will see that this generalization
does not lead to gauge-invariant answers unless we also include the
equations of motion corresponding to the magnetic 2-form potentials.
In this way we recover the observation of
\cite{deWit:2005ub,deWit:2008ta,deWit:2008gc,deWit:2007mt,Vroome:2007zd}
that magnetic gaugings require the introduction of magnetic 2-form
potentials in the action of the field theory.


\subsection{Purely electric gaugings}

Having $N=1,D=4$ supergravity in mind, we consider complex scalars
$Z^{i}\ (i=1,\cdots ,n)$ with K\"ahler metric $\mathcal{G}_{ij^{*}}$
admitting holomorphic Killing vectors
$K_{A}=k_{A}{}^{i}\partial_{i}+\mathrm{c.c.}$. The index $A$ of the
Killing vectors must be associated to those of the generators of the
global symmetry group $G$. In general, not all the global symmetries
will act on the scalars. Therefore, we assume that some of the
$K_{A}$ may be identically zero just as some of the matrices $T_{A\,
M}{}^{N}$ can be zero for other values of $A$. The action for the
electrically gauged theory is

\begin{equation}\label{actionel}
S_{\rm elec}[g,Z^{i},A^{\Lambda}]  =\int \left\{ \star R
-2\mathcal{G}_{ij^{*}}\mathfrak{D}Z^{i}
\wedge\star\mathfrak{D}Z^{*\, j^{*}} +2F^{\Sigma}\wedge G_{\Sigma}
-\star V \right\}\, ,
\end{equation}

\noindent
where $\mathfrak{D}Z^{i}$ is given by

\begin{equation}
\mathfrak{D}Z^{i}= dZ^{i} +A^{\Lambda}\Theta_{\Lambda}{}^{A}
k_{A}{}^{i}\, ,
\end{equation}

\noindent
and where $G_{\Lambda}$ denotes the combination of scalars and electric vector
field strengths defined by

\begin{equation}
G_{\Lambda}{}^{+}=f_{\Lambda\Sigma}(Z)F^{\Sigma\, +}\, ,
\end{equation}

\noindent where $F^{\Sigma\,+}=\tfrac{1}{2}(F^{\Sigma}+i\star
F^{\Sigma})$. It is assumed that the scalar-dependent kinetic matrix
$f_{\Lambda\Sigma}(Z)$ is invariant under the global symmetry group,
i.e.\footnote{Here we are only
  considering a restricted type of perturbative symmetries of the theory,
  excluding Peccei-Quinn-type shifts of the kinetic matrix for simplicity.
  We
  will consider these shifts together with the possible non-perturbative
  symmetries in the general gaugings' section.
  }

\begin{equation}
\label{eq:finvariance} \pounds_{A}f_{\Lambda\Sigma}  = 2T_{A\,
  (\Lambda}{}^{\Omega}f_{\Sigma) \Omega}\, ,
\end{equation}

\noindent
where $\pounds_{A}$ stands for the Lie derivative with respect to $K_A{}$,
since this is a pre-condition to gauge the theory.  However, the potential
needs only be invariant under the gauge transformations, because the gauging
usually adds to the globally-invariant potential of the ungauged theory
another piece.  Thus, we must have

\begin{equation}
\label{eq:Vinvariance} \pounds_{A}V = Y_{A\Lambda}{}^{C}\frac{\partial V}{\partial
  \Theta_{\Lambda}{}^{C}}\, ,
\end{equation}

\noindent
where $Y_{A\Lambda}{}^{C}$ is the electric component of the tensor defined in
Eq.~(\ref{eq:Ytensordef}). Indeed, using this property, one can show that
under the gauge transformations

\begin{equation}
  \begin{array}{rcl}
\delta Z^{i} & = & \Lambda^{\Lambda}\Theta_{\Lambda}{}^{A}k_{A}{}^{i}\, , \\
& & \\
\delta A^{\Lambda} & = & -\mathfrak{D}\Lambda^{\Lambda}\, ,\\
\end{array}
\end{equation}

\noindent
the scalar potential $V$ is gauge invariant:

\begin{equation}
\delta V =\Lambda^{\Sigma} \Theta_{\Sigma}{}^{A} \pounds_{A}V =
\Lambda^{\Sigma} Q_{\Sigma}{}^{\Lambda C}\frac{\partial V}{\partial
  \Theta_{\Lambda}{}^{A}}=0\, ,
\end{equation}

\noindent
on account of the quadratic constraint.

The equations of motion (plus the Bianchi identity for $F^{\Lambda}$)
corresponding to the action \eqref{actionel} are given by

\begin{eqnarray}
\mathcal{E}_{\mu\nu}
 & \equiv & -\star {\displaystyle\frac{\delta S}{\delta g^{\mu\nu}}} =
G_{\mu\nu} +2\mathcal{G}_{ij^{*}}[\mathfrak{D}_{\mu}Z^{i}
\mathfrak{D}_{\nu}Z^{*\, j^{*}} -{\textstyle\frac{1}{2}}g_{\mu\nu}
\mathfrak{D}_{\rho}Z^{i}\mathfrak{D}^{\rho}Z^{*\, j^{*}}]
\nonumber \\
& & \nonumber \\
& & -4\Im {\rm m}f_{\Lambda\Sigma} F^{\Lambda\,
+}{}_{\mu}{}^{\rho}F^{\Sigma\,
  -}{}_{\nu\rho}+{\textstyle\frac{1}{2}}g_{\mu\nu}V
\, ,
\\
& & \nonumber \\
\mathcal{E}_{i} & \equiv & {\textstyle \frac{1}{2}}
{\displaystyle\frac{\delta S}{\delta Z^{i}}} =
\mathcal{G}_{ij^{*}}\mathfrak{D}\star \mathfrak{D} Z^{*\, j^{*}}
-\partial_{i}G_{\Sigma}{}^{+}\wedge F^{\Sigma +} -\star
{\textstyle\frac{1}{2}}\partial_{i}V\, ,
\\
& & \nonumber \\
\mathcal{E}_{\Lambda} & \equiv &
 -{\textstyle\frac{1}{4}}{\displaystyle\star \frac{\delta S}{\delta  A^{\Lambda}}} =
\mathfrak{D}G_{\Lambda}-{\textstyle\frac{1}{4}}
\Theta_{\Lambda}{}^{A}\star j_{A}\, ,
\nonumber \\
& & \nonumber \\
\mathcal{E}^{\Lambda} & \equiv & \mathfrak{D}F^{\Lambda}\, ,
\end{eqnarray}

\noindent
where

\begin{equation}
j_{A} \equiv 2k^{*}_{A i}\mathfrak{D}Z^{i} +\mathrm{c.c.}\, ,
\end{equation}

\noindent
is the covariant Noether current.

According to the second Noether theorem there is an off-shell
relation between equations of motion of a theory associated to each
gauge invariance. For instance, associated to general covariance we
find the well-known identity

\begin{equation}
\nabla^{\mu}\mathcal{E}_{\mu\nu}
-(\mathfrak{D}_{\nu}Z^{i}\mathcal{E}^{*}_{i}+\mathrm{c.c.}) +2
F^{\Lambda}{}_{\nu\rho}(\star\mathcal{E}_{\Lambda})^{\rho}=0\, ,
\end{equation}

\noindent
which implies the on-shell covariant conservation of the energy-momentum
tensor.  Similarly, the identity associated to the Yang-Mills-type gauge
invariance of the theory is given by

\begin{equation}
\mathfrak{D}\mathcal{E}_{\Lambda}
+{\textstyle\frac{1}{2}}\Theta_{\Lambda}{}^{A} (k_{A}{}^{i}
\mathcal{E}_{i}+\mathrm{c.c.}) =0\, .
\end{equation}

\noindent
Using the Ricci identity for the covariant derivative and
Eqs.~(\ref{eq:finvariance}) and (\ref{eq:Vinvariance}) we find that this
equation is indeed satisfied because the Noether current satisfies the
identity

\begin{equation}
\label{Noetherid}
  \mathfrak{D}\star j_{A} = -2 (k_{A}{}^{i} \mathcal{E}_{i}+\mathrm{c.c.})
  +4 T_{A\, \Sigma}{}^{\Gamma}F^{\Sigma}\wedge G_{\Gamma}
  +\star Y_{A\Lambda}{}^{ C}\frac{\partial V}{\partial \Theta_{\Lambda}{}^{C}}\, .
\end{equation}

We are now going to establish a relation between the tensor hierarchy and the
equations of motion for the vector fields, their Bianchi identities and the
following projected scalar equations of motion:

\begin{eqnarray}
\mathfrak{D}G_{\Lambda}-{\textstyle\frac{1}{4}}
\Theta_{\Lambda}{}^{A}\star j_{A} & = & 0\, , \label{B1}\\
& & \nonumber \\
\mathfrak{D}F^{\Lambda} & = & 0 \, ,\label{B2}\\
& & \nonumber \\
\label{B3}
k_{A}{}^{i} \bigg [
\mathcal{G}_{ij^{*}}\mathfrak{D}\star \mathfrak{D} Z^{*\, j^{*}}
-\partial_{i}G_{\Sigma}{}^{+}\wedge F^{\Sigma +} -\star
{\textstyle\frac{1}{2}}\partial_{i}V\bigg ]
+\mathrm{c.c.} & = & 0\, .
\end{eqnarray}

\noindent
Note that, unlike the tensor hierarchy, these equations contain not only
$p$-form potentials but also the metric and scalars.

In order to derive the above equations of motion from the tensor hierarchy we
must complement the tensor hierarchy with a set of duality relations that
reproduces the scalar and metric dependence of these equations.  Besides the
usual $\mathfrak{D}^{2} Z$ term in the last equation the scalar dependence of
(\ref{B1})-(\ref{B3}) resides in the magnetic 2-forms $G_\Lambda$, the Noether
currents $j_A$ and the derivatives $\partial_i V$ of the scalar potential
$V$. The latter derivative is equivalently represented, via the invariance
property \eqref{eq:Vinvariance}, by the derivative
${\displaystyle\frac{\partial V}{\partial \Theta_{\Lambda}{}^{A}}}$ of the
scalar potential with respect to the embedding tensor. These are precisely the
objects that occur in the following set of duality relations that we
introduce:


\begin{equation}
  \begin{array}{rcl}
G_{\Lambda} & = & F_{\Lambda}\, ,
\\
& & \\
j_{A} & = & -2\star H_{A}\, ,\label{Dualityrules}\\
& &  \\
{\displaystyle\frac{\partial V}{\partial \Theta_{\Lambda}{}^{A}}} &
= & - 2\star G_{A}{}^{\Lambda}\, ,
\end{array}
\end{equation}

\noindent
where the magnetic 2-form field strengths $F_\Lambda$, the 3-form field
strengths $H_A$ and the 4-form field strengths $G_A{}^\Lambda$ are those of
the tensor hierarchy. The tensor hierarchy, together with the above duality
relations, forms the duality hierarchy. Upon hitting the duality relations
\eqref{Dualityrules} with a covariant derivative and next applying one of the
Bianchi identities of the tensor hierarchy we precisely obtain the equations
of motion (\ref{B1})-(\ref{B3}). In the case of the scalar equations of motion
we first obtain the identity

\begin{equation}
 \mathfrak{D}\star j_{A}
-4 T_{A\, \Sigma}{}^{\Gamma}F^{\Sigma}\wedge G_{\Gamma} -\star
Y_{A\Lambda}{}^{C}\frac{\partial V}{\partial
\Theta_{\Lambda}{}^{A}} =  0 \label{B4}\, .
\end{equation}

\noindent
Next, by comparing this equation with the Noether identity \eqref{Noetherid}
we derive the projected scalar equations of motion \eqref{B3}, i.e.

\begin{equation}
k_{A}{}^{i} \mathcal{E}_{i}+\mathrm{c.c.}  =  0.
\end{equation}

It also works the other way around. By substituting the duality
relations into the equations of motion the scalar and metric
dependence of these equations can be eliminated and one recovers the
hierarchy's Bianchi identities for a purely electric embedding
tensor $\Theta^{\Sigma A}=0$.  To be precise, Eqs.~(\ref{B1}) and
(\ref{B2}) are mapped into the 3-form Bianchi identities
(\ref{eq:bianchiFM}). Furthermore, Eq.~(\ref{B4}), which is
equivalent to \eqref{B3} upon use of the Noether identity
\eqref{Noetherid}, is mapped into the 4-form Bianchi identities
(\ref{eq:bianchiHA}).

We conclude that, at least in this case, the duality hierarchy encodes
precisely the vector equations of motion and the projected scalar equations of
motion via the duality rules (\ref{Dualityrules}).


\subsection{General gaugings}

In this subsection we wish to consider the more general case of electric and
magnetic gaugings. Our starting point is the formally symplectic-covariant
generalization of the equations of motion (\ref{B1})-(\ref{B3})\footnote{The
  Einstein and scalar equations of motion are just a rewriting of the original
  ones, which are already symplectic-invariant.}

\hskip -.5truecm\begin{eqnarray}
\mathcal{E}_{\mu\nu}
 & = &
G_{\mu\nu} +2\mathcal{G}_{ij^{*}}[\mathfrak{D}_{\mu}Z^{i}
\mathfrak{D}_{\nu}Z^{*\, j^{*}} -{\textstyle\frac{1}{2}}g_{\mu\nu}
\mathfrak{D}_{\rho}Z^{i}\mathfrak{D}^{\rho}Z^{*\, j^{*}}]
-G^{M}{}_{(\mu|}{}^{\rho}\star G_{M| \nu) \rho}
+{\textstyle\frac{1}{2}}g_{\mu\nu}V\, ,\nonumber
\\
& & \nonumber \\
\label{eq:Ei} \mathcal{E}_{i} & = &
\mathcal{G}_{ij^{*}}\mathfrak{D}\star \mathfrak{D} Z^{*\, j^{*}}
-\partial_{i}G_{M}{}^{+}\wedge G^{M +} -\star
{\textstyle\frac{1}{2}}\partial_{i}V\, ,
\\
& & \nonumber \\
\mathcal{E}_{M} & \equiv & \mathfrak{D}G_{M}-{\textstyle\frac{1}{4}}
\Theta_{M}{}^{A}\star j_{A}\, ,\nonumber
\end{eqnarray}

\noindent
where we have defined

\begin{equation}
(G^{M}) \equiv \left(
  \begin{array}{c}
  F^{\Sigma} \\ G_{\Sigma} \\
  \end{array}
\right)\, , \hspace{1cm}
 G_{\Sigma}{}^{+}= f_{\Sigma \Gamma}(Z) F^{\Gamma +}\, ,
\end{equation}

\noindent
and where the electric and magnetic field strengths $F^{M}$ are defined as in
the tensor hierarchy, i.e.~including the 2-form $B_{A}$ for which we do not
want to have an independent equation of motion to preserve the original number
of degrees of freedom.

The requirement that the kinetic matrix is invariant under the global symmetry
group $G$ and that the potential is gauge-invariant leads to the conditions

\begin{eqnarray}
\label{eq:finvariance2} \pounds_{A}f_{\Lambda\Sigma}  & = & -T_{A\,
\Lambda\Sigma} +2T_{A\, (\Lambda}{}^{\Omega}f_{\Sigma) \Omega}
-T_{A}{}^{\Omega\Gamma}f_{\Omega\Lambda}f_{\Gamma\Sigma}\, , \\
& & \nonumber \\
\label{eq:Vinvariance2} \pounds_{A}V & = &
Y_{AM}{}^{C}\frac{\partial V}{\partial  \Theta_{M}{}^{C}}\, ,
\end{eqnarray}

\noindent
from which it follows that

\begin{equation}
k_{A}{}^{i}\partial_{i} G_{M}{}^{+}\wedge G^{M +}
=k_{A}{}^{i}\partial_{i} f_{\Lambda\Sigma}F^{\Lambda +}\wedge
F^{\Sigma +} = -T_{A\, MN} G^{M}\wedge G^{N}\, .
\end{equation}

\noindent
A direct computation using the above properties leads to the following
identity for the covariant Noether current:

\begin{equation}\label{nc1}
\mathfrak{D}\star j_{A} = -2 (k_{A}{}^{i}
\mathcal{E}_{i}+\mathrm{c.c.}) -2 T_{A\, MN}G^{M}\wedge G^{N} +\star
Y_{A}{}^{\Lambda C}\frac{\partial V}{\partial
\Theta_{\Lambda}{}^{C}}\, .
\end{equation}

\noindent
On the other hand, the Ricci identity gives

\begin{equation}\label{r1}
\mathfrak{D}  \mathfrak{D} G_{M} = -X_{NM}{}^{P}F^{N}\wedge
G_{P}=X_{NPM}F^{N}\wedge G^{P}\, .
\end{equation}

\noindent
Taking the covariant derivative of the full $\mathcal{E}_{M}$ and using
Eqs.~(\ref{nc1}) and (\ref{r1}) we find

\begin{equation}
\mathfrak{D}\mathcal{E}_{M} +{\textstyle\frac{1}{2}}\Theta_{M}{}^{A}
(k_{A}{}^{i} \mathcal{E}_{i}+\mathrm{c.c.}) =
X_{NPM}(F^{N}-G^{N})\wedge G^{P} = \Theta^{\Sigma
A}(F_{\Sigma}-G_{\Sigma})\wedge T_{A\, PM}G^{P} \, .
\end{equation}

\noindent
This is the gauge identity associated to the standard electric and magnetic
gauge transformations of the vectors and scalars

\begin{equation}
  \begin{array}{rcl}
\delta Z^{i} & = & \Lambda^{M}\Theta_{M}{}^{A}k_{A}{}^{i}\, , \\
& & \\
\delta A^{M} & = & -\mathfrak{D}\Lambda^{M}\, ,\\
\end{array}
\end{equation}

\noindent provided that the right-hand side of the equation
vanishes. Since this is not the case we conclude that the equations
of motion are not gauge-invariant. Hence, a naive symplectic
covariantization of the electric gauging case is not enough to
obtain a gauge-invariant answer involving magnetic gaugings.

In order to re-obtain gauge invariance
we extend the set of equations of motion, adding, arbitrarily, as equation of
motion of the 2-forms $B_{A}$

\begin{equation}
\mathcal{E}^{A}\equiv
\Theta^{MA}(F_{M}-G_{M})
=
-\Theta^{\Sigma A}(F_{\Sigma}-G_{\Sigma})\, ,
\end{equation}

\noindent
so that the above identity becomes again a relation between equations of
motion

\begin{equation}
\mathfrak{D}\mathcal{E}_{M} +{\textstyle\frac{1}{2}}\Theta_{M}{}^{A}
(k_{A}{}^{i} \mathcal{E}_{i}+\mathrm{c.c.}) + T_{A\,
MP}\mathcal{E}^{A}\wedge G^{P}=0 \, ,
\end{equation}

\noindent
that we can interpret as the gauge identity associated to an off-shell gauge
invariance of the extended set of equations of motion.

The price we may have to pay for doing this is the possible modification of
the equations of motion of the vector fields: the above gauge identities are
associated to the gauge transformations of $B_{A}$

\begin{equation}
\delta B_{A}= 2T_{A\, MP}\Lambda^{M}G^{P} +2R_{A\, M} \wedge \delta
A^{M}\, ,
\end{equation}

\noindent
where $R_{A\, M}$ is a 1-form that is cancelled in the above gauge identity by
an extra term in the equation of motion of the vector fields:

\begin{equation}
\mathcal{E}^{\prime}_{M}= \mathcal{E}_{M} +R_{A\, M} \mathcal{E}^{A}
\wedge A^{M}\, .
\end{equation}

\noindent
The 1-forms $R_{A\, M}$ must be such that the infinitesimal gauge
transformations form a closed algebra. The gauge identity takes now the form

\begin{equation}
\mathfrak{D}\mathcal{E}^{\prime}_{M}
+{\textstyle\frac{1}{2}}\Theta_{M}{}^{A} (k_{A}{}^{i}
\mathcal{E}_{i}+\mathrm{c.c.}) + T_{A\, MP}\mathcal{E}^{A}\wedge
G^{P} -\mathfrak{D}(R_{A\, M} \mathcal{E}^{A} \wedge A^{M})=0 \, .
\end{equation}

\noindent
In order to make contact with the tensor hierarchy we take $R_{A\,
  M}={\textstyle\frac{1}{2}}X^{P}{}_{MN}A^{N}\wedge (F_{P}-G_{P})$.

We observe that the equations of motion also satisfy the relation

\begin{equation}
\mathfrak{D}\mathcal{E}^{A} -{\textstyle\frac{1}{2}}T_{B\,
MN}\Theta^{PA}A^{N} \wedge \mathcal{E}^{B}
+\Theta^{MA}\mathcal{E}_{M}=0\, ,
\end{equation}

\noindent
which can be interpreted as the gauge identity associated to the symmetry

\begin{equation}
\begin{array}{rcl}
  \delta A^{M} & = & Z^{MA}\Lambda_{A}\, ,\\
  & & \\
  \delta B_{A} & = & \mathfrak{D}\Lambda_{A}
  -{\textstyle\frac{1}{2}}T_{AMN}\Theta^{NB} A^{M}\wedge \Lambda_{B}\, .\\
\end{array}
\end{equation}

As we did in the electric gauging case, we are now going to establish a
relation between the tensor hierarchy and the following equations of motion:

\begin{eqnarray}
\mathcal{E}^{\prime}_{M} & = &
\mathfrak{D}G_{M}-{\textstyle\frac{1}{4}} \Theta_{M}{}^{A}\star
j_{A}
+{\textstyle\frac{1}{2}}T_{A\, MN}A^{N}\wedge \Theta^{PA}(F_{P}-G_{P})=0 \, , \\
& & \nonumber \\
\mathcal{E}^{A} & = & \Theta^{MA}(F_{M}-G_{M}) =0\, ,\\
& & \nonumber \\
k_{A}{}^{i}\mathcal{E}_{i} & = & k_{A}{}^{i}\bigg [ \mathcal{G}_{ij^{*}}\mathfrak{D}\star \mathfrak{D} Z^{*\, j^{*}}
-\partial_{i}G_{M}{}^{+}\wedge G^{M +} -\star
{\textstyle\frac{1}{2}}\partial_{i}V\bigg ] = 0\,.
\end{eqnarray}

\noindent
These equations are invariant under the gauge transformations

\begin{eqnarray}
\label{eq:actiongaugetransformations1} \delta_{a}Z^{i} & = &
\delta_{h}Z^{i}\, ,
\\
& & \nonumber \\
\label{eq:actiongaugetransformations2} \delta_{a} A^{M} & = &
\delta_{h} A^{M}\, ,
\\
& & \nonumber \\
\label{eq:actiongaugetransformations3} \delta_{a} B_{A} & = &
\delta_{h}B_{A} -2T_{A\,
  NP}\Lambda^{N}(F^{P}-G^{P})\, ,
\end{eqnarray}

\noindent
where we have denoted by $\delta_{a}$ the gauge transformations that leave
this system of equations invariant and by $\delta_{h}$ those derived in the
construction of the 4-dimensional tensor hierarchy (summarized in
Appendix~\ref{app-gaugetranshierarchy}). $\delta_{a} B_{A}$ is, therefore,
just $\delta_{h} B_{A}$ with $F^{P}$ replaced by $G^{P}$.

Following the electric gauging case, in order to derive the above equations of
motion from the tensor hierarchy, we introduce the following set of duality
relations:

\begin{equation}
\label{eq:map}
  \begin{array}{rcl}
G^{M} & = & F^{M}\, ,
\\
& & \\
j_{A} & =& -2\star H_{A}\, ,\\
& &  \\
{\displaystyle\frac{\partial V}{\partial \Theta_{M}{}^{A}}} & = & -
2\star G_{A}{}^{M}\, .
\end{array}
\end{equation}

\noindent We note that the gauge-covariance of the first duality
relation is more subtle in that $G^{M}$ transforms not only
covariantly, but also into $G^{M}-F^{M}$, see \cite{DeRydt:2008hw}.
Note that the equation of motion of the magnetic 2-form potentials,
$\mathcal{E}^{A}=0$, is identified as a projected duality relation.
To recover the other equations of motion we have to again hit the
duality relations \eqref{eq:map} with a covariant derivative and
next apply one of the Bianchi identities of the tensor hierarchy. To
derive the projected scalar equations of motion we first obtain the
identity

\begin{equation}
\mathfrak{D}\star j_{A} +2 T_{A\, MN}G^{M}\wedge G^{N} -\star
Y_{A}{}^{\Lambda C}\frac{\partial V}{\partial
  \Theta_{\Lambda}{}^{A}} \ = \ 0
\end{equation}

\noindent
from the duality hierarchy and, next, apply the Noether identity \eqref{nc1}.


The gauge identities guarantee the existence of a gauge-invariant action from
which the equations of motion $\mathcal{E}_{M}^{\prime}$ and $\mathcal{E}^{A}$
can be derived. This action has actually been constructed in
Ref.~\cite{deWit:2005ub}.  In our conventions, it is given by

\begin{equation}
\label{eq:action1-2-forms}
\begin{array}{rcl}
S[g_{\mu\nu},Z^{i},A^{M},B_{A}]  & = &  {\displaystyle\int} \left\{
\star R -2\mathcal{G}_{ij^{*}}\mathfrak{D}Z^{i}
\wedge\star\mathfrak{D}Z^{*\, j^{*}} +2F^{\Sigma}\wedge G_{\Sigma}
-\star V \right.
\\[-0.1cm]
& & \\
& & -4Z^{\Sigma A}B_{A}\wedge \left(F_{\Sigma}
-{\textstyle\frac{1}{2}}Z_{\Sigma}{}^{B}B_{B}\right)
\\
& & \\
& &
-{\textstyle\frac{4}{3}}  X_{[MN]\Sigma} A^{M}\wedge A^{N} \wedge
\left(F^{\Sigma} -Z^{\Sigma B}B_{B}\right)
\\
& & \\
& & \left. -{\textstyle\frac{2}{3}}  X_{[MN]}{}^{\Sigma}A^{M}\wedge
A^{N} \wedge \left(dA_{\Sigma} -{\textstyle\frac{1}{4}}
X_{[PQ]\Sigma} A^{P}\wedge A^{Q}\right)
\right\}\, .\\
\end{array}
\end{equation}

\noindent
A general variation of the above action gives

\begin{equation}
\label{eq:variationaction1-2-forms}
\delta S
=
{\displaystyle\int}
\left\{
\delta g^{\mu\nu}  {\displaystyle\frac{\delta S}{\delta
g^{\mu\nu}}}
+\left(\delta Z^{i}  {\displaystyle\frac{\delta S}{\delta Z^{i}}}
+\mathrm{c.c.}\right)
-\delta A^{M}\wedge \star{\displaystyle\frac{\delta S}{\delta A^{M}}}
+2\delta B_{A} \wedge\star {\displaystyle\frac{\delta S}{\delta  B_{A}}}
\right\}\, ,
\end{equation}

\noindent
where

\begin{eqnarray}
\label{eq:SEmn}
{\displaystyle\frac{\delta S}{\delta g^{\mu\nu}}}
& = &
\star \mathbb{I}\mathcal{E}_{\mu\nu}\, ,\\
& & \nonumber \\
\label{eq:SEi} -{\textstyle \frac{1}{2}} {\displaystyle\frac{\delta
S}{\delta Z^{i}}} & = &
\mathcal{E}_{i}\, ,\\
& & \nonumber \\
-{\textstyle\frac{1}{4}}{\displaystyle\star \frac{\delta S}{\delta
A^{M}}} & = &
\mathcal{E}^{\prime}_{M}\, ,\\
& & \nonumber \\
{\displaystyle\star \frac{\delta S}{\delta  B_{A}}} & = &
\mathcal{E}^{A}\, .
\end{eqnarray}


\subsection{The unconstrained case}

In this subsection we briefly comment on the meaning of the top-form
and next to top-form potentials. Experience shows that these
higher-rank potentials can be related to constraints: the constancy
of $\Theta_{M}{}^{A}$, $\mathfrak{D}\Theta_{M}{}^{A}=0$, can be
associated to the 3-form potential, and the quadratic and linear
constraints $Q_{NP}{}^{E}=0$, $Q^{AB}=0$, $L_{NPQ}=0$ can be
associated to the 4-form potentials $D_{E}{}^{NP}, D_{AB}, D^{NPQ}$
that we have providentially found. We would like to stress, however,
that prior to relaxing the constraints one is forced to introduce
these potentials if one requires that the field equations are
derivable as compatibility conditions from the duality relations, as
we showed in the previous section.

In view of the discussion of an action principle with Lagrange
multipliers in the next section, we reconsider the gauge identities
of the equations $\mathcal{E}^{\prime}_{M},\mathcal{E}^{A}$ defined
in the previous subsections assuming that those constraints are not
satisfied. We then denote the embedding tensor by
$\vartheta_{M}{}^{A}=\vartheta_{M}{}^{A}(x)$ in order to indicate
that it is now space-time dependent. Evidently, we are going to get
extra terms proportional to the constraints which we will
reinterpret as equations of motion of the 3- and 4-form potentials,
obtaining new gauge identities that involve the equations of motion
of all fields. Thus, off-shell gauge invariance will have been
preserved by the same mechanism used in the previous case. The price
that we will have to pay is the same: modifying the gauge
transformations and the equations of motion.

This procedure is too complicated in this case, though. As an example, let us
take the covariant derivative of $\mathcal{E}^{A}$:

\begin{equation}
\mathfrak{D}\mathcal{E}^{A}
=
-\mathfrak{D}\vartheta_{M}{}^{A}\wedge(F^{M}-G^{M})
+\vartheta^{MA}(\mathfrak{D}F_{M}-\mathfrak{D}G_{M})\, .
\end{equation}

\noindent
The unconstrained Bianchi identity for $F^{M}$ is

\begin{equation}
\label{eq:unconstrainedbianchiFM}
\begin{array}{rcl}
\mathfrak{D}F^{M} & = & Z^{MB}[H_{B} -Y_{BN}{}^{C}C_{C}{}^{N}]
+L^{M}{}_{RS}[{\textstyle\frac{3}{2}}A^{R}\wedge dA^{S}
+{\textstyle\frac{1}{2}}X_{NP}{}^{S}A^{R}\wedge A^{N}\wedge A^{P} ] \\
& & \\
& & +\mathfrak{D}\vartheta_{N}{}^{A} \wedge [{\textstyle\frac{1}{2}}
\Omega^{NM}B_{A} +{\textstyle\frac{1}{2}}T_{A\, P}{}^{M}A^{N}\wedge
A^{P}] +{\textstyle\frac{1}{3}}Q_{NP}{}^{E}T_{E\, R}{}^{M} A^{N}
\wedge
A^{P} \wedge A^{R} \, , \\
\end{array}
\end{equation}

\noindent
and, using the equation of motion $\mathcal{E}^{\prime}_{M}$ we can write the
following gauge identity

\begin{equation}
\begin{array}{rcl}
\mathfrak{D}\mathcal{E}^{A} -{\textstyle\frac{1}{2}}T_{B\,
MN}\vartheta^{MA}A^{N} \wedge \mathcal{E}^{B}
+\vartheta^{MA}\mathcal{E}^{\prime}_{M}
+Q^{AB}[2(H_{B}+{\textstyle\frac{1}{2}}\star j_{B}) -2
Y_{BN}{}^{C}C_{C}{}^{N}]
& & \\
& & \\
+\mathfrak{D}\vartheta_{M}{}^{B} \wedge [{\textstyle\frac{1}{2}}
\vartheta^{MA}B_{B} +{\textstyle\frac{1}{2}} T_{B\,
P}{}^{Q}\vartheta_{Q}{}^{A}A^{M}\wedge A^{P} +\delta_{B}{}^{A}
(F^{M}-G^{M})]
& & \\
& & \\
+L_{MRS}\vartheta^{MA} [-{\textstyle\frac{3}{2}}A^{R}\wedge dA^{S}
-{\textstyle\frac{1}{2}}X_{NP}{}^{S} A^{R}\wedge A^{N} \wedge A^{P}]
& & \\
& & \\-{\textstyle\frac{1}{3}}Q_{NP}{}^{E}T_{E\, R}{}^{M}
\vartheta_{M}{}^{A} A^{N} \wedge A^{P} \wedge A^{R}
& = & 0\, .\\
\end{array}
\end{equation}

\noindent
It is very difficult to infer directly from this and similar identities all
the gauge transformations of the fields and the modifications of the equations
of motion. Thus, we are going to adopt a different strategy in the next
section: we are going to construct directly a gauge-invariant action.



\section{The $D=4$ action}
\label{sec-4daction}

In this section we perform the third and last step of our procedure:
the construction of an action for the fields of the tensor
hierarchy\footnote{Actually, not all the 2-forms $B_A$ will appear
in the
  action but only $\Theta^{\Lambda A}B_A$.}. Our starting point is the action
Eq.~(\ref{eq:action1-2-forms}), which we will denote by $S_{0}$ in
what follows and which includes, besides the metric, only scalars,
1-forms and 2-forms and which is invariant under the gauge
transformations
Eqs.~(\ref{eq:actiongaugetransformations1})-(\ref{eq:actiongaugetransformations3}).
We now want to add to it 3- and 4-forms as Lagrange multipliers
enforcing the covariant constancy of the embedding tensor (which we
promote to an unconstrained scalar field $\Theta_{M}{}^{A}(x)$)
and the three algebraic constraints $Q^{AB}$, $L_{NPQ}$,
$Q_{NP}{}^{E}$ that we have imposed on the embedding tensor. The new
terms must be metric-independent (``topological'') and
scalar-independent in order to leave unmodified the scalar and
Einstein equations of motion
(\ref{eq:Ei}) which are derived from the action $S_{0}$ given in
Eq.~(\ref{eq:action1-2-forms}).

Thus, we add to $S_{0}$ the following piece $\Delta S$ given
by\footnote{Observe that $\mathfrak{D}\Theta_{M}{}^{A}=d\Theta_{M}{}^{A}
  -Q_{NM}{}^{A}A^{N}$ and, therefore, the covariant constancy of the embedding
  tensor plus the quadratic constraint $Q_{NP}{}^{E}=0$ imply
  $d\Theta_{M}{}^{A}=0$.}

\begin{equation}
\Delta S
=
\int \left\{
 \mathfrak{D}\vartheta_{M}{}^{A}\wedge \tilde{C}_{A}{}^{M}
+Q_{NP}{}^{E}\tilde{D}_{E}{}^{NP} +Q^{AB}\tilde{D}_{AB}
+L_{NPQ}\tilde{D}^{NPQ} \right\}\, .
\end{equation}

\noindent
The tildes in $\tilde{C}_{C}{}^{M}$, $\tilde{D}_{AB}$, $\tilde{D}^{NPQ}$ and
$\tilde{D}_{E}{}^{NP}$ indicate that these 3- and 4-form fields need not be
identical to those found in the hierarchy, although we expect them to be
related by field redefinitions.

The action $S_{0}$ is no longer gauge invariant under the
gauge transformations involving 0- and 1-form gauge parameters
$\Lambda^{M},\Lambda_{A}$, without imposing any constraints on the embedding tensor,
but the non-vanishing terms in the transformation
can only be proportional to the l.h.s.'s of the constraints
$\mathfrak{D}\vartheta_{M}{}^{C}=0$, $Q_{NP}{}^{E}=0$, $Q^{AB}=0$ and
$L_{NPQ}=0$ and, by choosing appropriately the gauge transformations of
$\tilde{C}_{C}{}^{M}$, $\tilde{D}_{AB}$, $\tilde{D}^{NPQ}$ and
$\tilde{D}_{E}{}^{NP}$ we can always make the variation of the action $S\equiv S_0+\Delta S$
vanish. Having done that we would like to relate the tilded fields with the
untilded ones in the hierarchy.

Let us start by computing the general variation of the action.
Taking into account the fact that the fields $g_{\mu\nu}$, $Z^{i}$
and $B_{A\, \mu\nu}$ only occur in $S_{0}$, that the field
$A^{M}{}_{\mu}$ occurs in $S_0$ and in the term
$\mathfrak{D}\vartheta_{M}{}^{A}\tilde{C}_{A}{}^{M}$ in $\Delta S$
and that the new fields $\tilde{C}_{C}{}^{M}$, $\tilde{D}_{AB}$,
$\tilde{D}^{NPQ}$ and $\tilde{D}_{E}{}^{NP}$ only occur in $\Delta
S$, we find

\begin{equation}
\label{eq:variationtotalaction}
\begin{array}{rcl}
\delta S & = & {\displaystyle\int} \left\{ \delta g^{\mu\nu}
{\displaystyle\frac{\delta S_{0}}{\delta g^{\mu\nu}}} +\left(\delta
Z^{i}  {\displaystyle\frac{\delta S_{0}}{\delta Z^{i}}}
+\mathrm{c.c.}\right) -\delta A^{M}\wedge \star
{\displaystyle\frac{\delta S_{0}}{\delta A^{M}}} +2\delta B_{A}
\wedge \star {\displaystyle\frac{\delta S_{0}}{\delta  B_{A}}}
\right.
\\
& & \\
& & +\mathfrak{D}\vartheta_{M}{}^{A}\wedge \delta
\tilde{C}_{A}{}^{M} +Q_{NP}{}^{E} (\delta
\tilde{D}_{E}{}^{NP}-\delta A^{N}\wedge \tilde{C}_{E}{}^{P})
+Q^{AB}\delta \tilde{D}_{AB}
\\
& & \\
& & \left. +L_{NPQ} \delta \tilde{D}^{NPQ} +\delta
\vartheta_{M}{}^{A} {\displaystyle\frac{\delta S}{\delta
\vartheta_{M}{}^{A}}} \right\}\, .
\end{array}
\end{equation}

The scalar and Einstein equations of motion are as in Eqs.~(\ref{eq:Ei}) and
(\ref{eq:SEmn}),(\ref{eq:SEi}). The variations of the old action $S_{0}$ with
respect to $A^{M}$ and $B_{A}$ are modified by terms proportional to the
constraints. We can write them in the form

\begin{eqnarray}
-{\textstyle\frac{1}{4}}{\displaystyle\star \frac{\delta
S_{0}}{\delta  A^{M}}} & = &
\mathfrak{D}F_{M}-{\textstyle\frac{1}{4}}\vartheta_{M}{}^{A}\star
j_{A} -{\textstyle\frac{1}{3}}dX_{[PQ] M}\wedge A^{P}\wedge A^{Q}
-{\textstyle\frac{1}{2}}Q_{(NM)}{}^{E}A^{N}\wedge B_{E}
\nonumber \\
& & \nonumber \\
& & -L_{MNP}A^{N}\wedge \left(dA^{P}
+{\textstyle\frac{3}{8}}X_{[RS]}{}^{P}A^{R}\wedge A^{S}\right)
+{\textstyle\frac{1}{8}}Q_{NP}{}^{A}T_{A\, QM}A^{N} \wedge A^{P}
\wedge A^{Q}
\nonumber \\
& & \nonumber \\
& & -d(F_{M}-G_{M}) -X_{[MN]}{}^{P}A^{N}\wedge(F_{P}-G_{P}) \, ,
\label{eq:EM}
\\
& & \nonumber \\
{\displaystyle\star \frac{\delta S_{0}}{\delta  B_{A}}} & = &
\vartheta^{P A}(F_{P}-G_{P}) +Q^{AB}B_{B}\, . \label{eq:EA}
\end{eqnarray}

\noindent
In deriving these equations we have used the unconstrained Bianchi identity
for $F^{\Lambda}$, given by the upper component of
Eq.~(\ref{eq:unconstrainedbianchiFM}), to replace $H_{A}$ in the equation of
motion of $A_{\Lambda}$.  This has allowed us to write a
symplectic-covariant expression for the equation of motion of $A^{M}$.

The only non-trivial variation that remains to be computed in
Eq.~(\ref{eq:variationtotalaction}) is the equation of motion of the embedding
tensor. We get

\begin{equation}
  \begin{array}{rcl}
{\displaystyle \frac{\delta S}{\delta  \vartheta_{M}{}^{A}}} & = &
-\mathfrak{D}\tilde{C}_{A}{}^{M} +Z^{MB}B_{B}\wedge B_{A}-2
(F^{M}-G^{M})\wedge B_{A} -\star {\displaystyle\frac{\partial
V}{\partial
    \vartheta_{M}{}^{A}}}
\\
& & \\
& & +W_{ANP}{}^{EM} \tilde{D}_{E}{}^{NP} +W_{A}{}^{BCM}
\tilde{D}_{BC} +W_{ANPQ}{}^{M} \tilde{D}^{NPQ}
\\
& & \\
& & +A^{M}\wedge \left\{ -\star j_{A}
+Y_{AN}{}^{C}\tilde{C}_{C}{}^{N}
-T_{AN}{}^{P}A^{N}\wedge(F_{P}-G_{P}) \right.
\\
& & \\
& & \left. -\frac{4}{3}T_{ANR}A^{N}\wedge \left[ dA^{R}
+\frac{3}{8}X_{[PQ]}{}^{R}A^{P}\wedge A^{Q}
 +\frac{3}{2}Z^{RB}B_{B}
\right] \right\}\, .
  \end{array}
\end{equation}

We are going to use this equation to find the relation between the tilded
fields and the hierarchy fields. Using Eqs.~(\ref{eq:map}) and the definitions
of the tensor hierarchy's field strengths $H_{A}$ and $G_{A}{}^{M}$, we are
left with

\begin{equation}
  \begin{array}{rcl}
{\textstyle\frac{1}{2}} {\displaystyle \frac{\delta S}{\delta
\vartheta_{M}{}^{A}}} & = & \mathfrak{D}(-{\textstyle\frac{1}{2}}
\tilde{C}_{A}{}^{M} -C_{A}{}^{M} -A^{M}\wedge B_{A})
\\
& & \\
& & +Y_{AP}{}^{C}A^{M}\wedge ({\textstyle\frac{1}{2}}
\tilde{C}_{C}{}^{P} +C_{C}{}^{P} +A^{P}\wedge B_{C}) +W_{A}{}^{BCM}
({\textstyle\frac{1}{2}}\tilde{D}_{BC}-D_{BC})
\\
& & \\
& & +W_{ANP}{}^{EM}
({\textstyle\frac{1}{2}}\tilde{D}_{E}{}^{NP}-D_{E}{}^{NP}
+{\textstyle\frac{1}{2}}A^{N}\wedge A^{P} \wedge B_{E})
\\
& & \\
& & +W_{ANPQ}{}^{M} ({\textstyle\frac{1}{2}}\tilde{D}^{NPQ}-
D^{NPQ})\, ,
  \end{array}
\end{equation}

\noindent
which is satisfied if we identify

\begin{equation}
\label{eq:34formidentification}
  \begin{array}{rclrcl}
\tilde{C}_{A}{}^{M} & = & -2(C_{A}{}^{M}+A^{M}\wedge B_{A})\,
,\hspace{1cm} & \tilde{D}_{E}{}^{NP} & = & 2D_{E}{}^{NP}-A^{N}\wedge
A^{P} \wedge B_{E}\, ,
\\
& & & & & \\
\tilde{D}_{BC} & = & 2 D_{BC}\, , &
\tilde{D}^{NPQ} & = & 2D^{NPQ}\, .\\
\end{array}
\end{equation}

\noindent
Using these identifications $\Delta S$ reads

\begin{equation}
\label{eq:DeltaS}
\begin{array}{rcl}
\Delta S  & = &  {\displaystyle\int} \left\{
-2\mathfrak{D}\vartheta_{M}{}^{A}\wedge (C_{A}{}^{M} +A^{M}\wedge
B_{A}) +2Q_{NP}{}^{E}(D_{E}{}^{NP}
-{\textstyle\frac{1}{2}}A^{N}\wedge A^{P} \wedge B_{E}) \right.
\\
& & \\
& & \left. +2Q^{AB}D_{AB} +2L_{NPQ}D^{NPQ}
\right\}\, ,\\
\end{array}
\end{equation}

\noindent
and a general variation of the total action $S=S_0+\Delta S$ is given by

\begin{equation}
\begin{array}{rcl}
\delta S & = & {\displaystyle\int} \left\{ \delta g^{\mu\nu}
{\displaystyle\frac{\delta S_{0}}{\delta g^{\mu\nu}}} +\left(\delta
Z^{i} {\displaystyle\frac{\delta S_{0}}{\delta Z^{i}}}
+\mathrm{c.c.}\right) -\delta A^{M}\wedge \star
{\displaystyle\frac{\delta S_{0}}{\delta A^{M}}} +2\delta B_{A}
\wedge \star {\displaystyle\frac{\delta S_{0}}{\delta  B_{A}}}
\right.
\\
& & \\
& & +\mathfrak{D}\vartheta_{M}{}^{A}\wedge [-2\delta  C_{A}{}^{M}
-2\delta A^{M} \wedge B_{A} -2 A^{M}\wedge \delta B_{A} ]
+Q^{AB}[2\delta D_{AB}]
\\
& & \\
& & +Q_{NP}{}^{E}[ 2\delta D_{E}{}^{NP} +2 \delta A^{N} \wedge
C_{E}{}^{P} +2\delta A^{(N}\wedge A^{P)} \wedge B_{E} -A^{N}\wedge
A^{P} \wedge \delta B_{E}]
\\
& & \\
& & \left. +L_{NPQ}[2 \delta D^{NPQ}] +\delta \vartheta_{M}{}^{A}
{\displaystyle\frac{\delta  S}{\delta  \vartheta_{M}{}^{A}}}
\right\}\, .
\end{array}
\end{equation}

\noindent
The first variation of the total action $S$ with respect to
$\vartheta_{M}{}^{A}$ can be written in the form

\begin{equation}
  \begin{array}{rcl}
{\textstyle\frac{1}{2}} {\displaystyle \frac{\delta S}{\delta
\vartheta_{M}{}^{A}}} & = & (G_{A}{}^{M}
-{\textstyle\frac{1}{2}}\star\partial V/\partial \vartheta_{M}{}^{A}
) -A^{M}\wedge (H_{A}+{\textstyle\frac{1}{2}}\star j_{A})
\\
& & \\
& & -{\textstyle\frac{1}{2}}T_{AN}{}^{P}A^{M}\wedge A^{N}\wedge
(F_{P}-G_{P}) -(F^{M}-G^{M})\wedge B_{A}\, . \label{eq:EAM}
\end{array}
\end{equation}

We can now check the gauge invariance of the total action $S$. We are going to
use for the gauge transformations of all the fields (except for the scalars
and vectors) the Ansatz $\delta_{a}=\delta_{h}+\Delta$ where $\Delta$ is a
piece to be determined. If we assume that the embedding tensor is exactly
invariant\footnote{One could also allow $\vartheta_M^A$ to transform according
  to its indices as $\delta\vartheta_M^A=-Q_{NM}{}^A\Lambda^N$. This is like
  adding a term proportional to an equation of motion, that of $D_A{}^{NM}$,
  to the zero variation.}, i.e.  $\delta \vartheta_{M}{}^{A} = 0$, we find

\begin{eqnarray}
\Delta B_{A} & = & -2T_{A\, NP}\Lambda^{N}(F^{P}-G^{P})\, ,
\\
& & \nonumber \\
\Delta C_{A}{}^{M} & = & \Lambda_{A}\wedge  (F^M-G^M)
-\Lambda^{M}(H_{A}+{\textstyle\frac{1}{2}}\star j_{A})\, ,
\\
& & \nonumber \\
\Delta D_{AB} & = & 2\Lambda_{[A}\wedge
(H_{B]}+{\textstyle\frac{1}{2}}\star j_{B]}) -2T_{[A|\,
NP}\Lambda^{N}(F^{P}-G^{P})\wedge B_{|B]}
\, , \\
& & \nonumber \\
\Delta D_{E}{}^{NP} & = & -\Lambda^{N}(G_{E}{}^{P}
-{\textstyle\frac{1}{2}}\star\partial V/\partial \vartheta_{P}{}^{E}
) +(F^{N}-G^{N})\wedge \Lambda_{E}{}^{P}\, ,
\\
& &  \nonumber \\
\Delta D^{NPQ} & = & -3\delta A^{(N}\wedge A^{P} \wedge
(F^{Q)}-G^{Q)}) +6 \Lambda^{(N}F^{P}\wedge  (F^{Q)}-G^{Q)})
 \nonumber  \\
& & \nonumber \\
& & -3 \Lambda^{(N}  (F^{P}-G^{P})\wedge  (F^{Q)}-G^{Q)})\, ,
\end{eqnarray}

\noindent
where we have used in this calculation the non-trivial Ricci
identities\footnote{If the constraints are satisfied, $\vartheta_{M}{}^{C}
  \mathfrak{D} \mathfrak{D} \Lambda_{C}{}^{M}= \mathfrak{D} \mathfrak{D}
  (\vartheta_{M}{}^{C}\Lambda_{C}{}^{M})=dd
  (\vartheta_{M}{}^{C}\Lambda_{C}{}^{M})=0$. Therefore, when they are not
  satisfied, $\vartheta_{M}{}^{C} \mathfrak{D} \mathfrak{D} \Lambda_{C}{}^{M}$
  must be proportional to them.}

\begin{eqnarray}
\vartheta_{M}{}^{C} \mathfrak{D} \mathfrak{D} \Lambda_{C}{}^{M} & =
& \mathfrak{D}\vartheta_{M}{}^{A}\wedge (-Y_{AP}{}^{E}A^{M}\wedge
\Lambda_{E}{}^{P}) +Q_{NP}{}^{E} [(F^{N}-Z^{NA}B_{A})\wedge
\Lambda_{E}{}^{P}
\nonumber \\
& & \nonumber \\
& & -{\textstyle\frac{1}{2}}Y_{EQ}{}^{C}A^{N}\wedge A^{P}\wedge
\Lambda_{C}{}^{Q}]\, , \\
& & \nonumber \\
\mathfrak{D}\mathfrak{D}F_{M} & = & X_{NPM}F^{N}\wedge F^{P} -2
Q^{AB}T_{A\, PM}F^{P}\wedge B_{B} +dX_{NPM}\wedge A^{N}\wedge F^{P}
\nonumber \\
& & \nonumber \\
& & -{\textstyle\frac{1}{2}} Q_{NP}{}^{E} T_{E\, MQ}  A^{N}\wedge
A^{P} \wedge F^{Q}\, ,
\end{eqnarray}

\noindent
and the variations of the kinetic matrix and the potential
Eqs.~(\ref{eq:finvariance2}) and (\ref{eq:Vinvariance2}).

We observe that all terms in the extra variations $\Delta$ vanish when we use
the duality relations (\ref{eq:map}). Actually, all of them, except for just
one term in $\Delta D^{NPQ}$, are such that the variations $\delta_a$ are
obtained from the tensor hierarchy variations $\delta_{h}$ simply by replacing
the scalar-independent field strengths $F^M, H_A, G_A{}^M$ by the
corresponding scalar-dependent objects $G^M, j_A, {\displaystyle\frac{\partial
    V}{\partial \vartheta_{\Lambda}{}^{A}}}$ via the duality relations
(\ref{eq:map}).

Finally, we note that the variations $\delta_a$ and $\delta_h$ are equivalent
from the point of view of the duality hierarchy. The two sets of transformation
rules differ by terms that are proportional to the duality
relations. The only difference is that the commutator algebra corresponding to $\delta_h$ closes off-shell whereas the
algebra corresponding to $\delta_a$ closes up to terms that are
proportional to the duality relations.
The two sets of transformation rules are not equivalent from the
action point of view in the sense that only one of them, the one with transformation rules $\delta_a$,
leaves the action invariant, whereas the other, with transformations $\delta_h$, does not.


\section{The 3-dimensional case}
\label{sec-3d}

As an illustration of our general procedure we will construct in
this section the complete $D=3$ tensor and duality hierarchy
corresponding to a generic $D=3$ gauged supergravity theory,
extending the analysis of the maximally supersymmetric case
\cite{deWit:2008ta,Bergshoeff:2008qd}. The $D=3$ hierarchy is
sufficiently short in order to allow for a straightforward analysis
and nevertheless captures the features expected to appear in general
dimensions.


\subsection{Generalities on $D=3$}
\label{MaximalD=3}

Three-dimensional gauged supergravity has been constructed in
\cite{Nicolai:2000sc,Nicolai:2001sv} for the maximal case and subsequently
generalized to lower supersymmetries in \cite{Nicolai:2001ac,deWit:2003ja}.

$D=3$ (ungauged) supergravities are particularly simple theories because their
only physical bosonic degrees of freedom are described by scalar fields, since
in $D=3$ the metric and $p$-forms with $p\geq 2$ have no dynamics and vectors
are dual to scalars. The number of scalar fields as well as the rigid symmetry
group $G$ is ultimately constrained by supersymmetry. For instance, in case of
maximal supersymmetry there are $128$ scalars, which parameterize the coset
space $E_{8(8)}/SO(16)$, and thus we have $G=E_{8(8)}$.  However, for the
general construction of the tensor hierarchy to be discussed here
supersymmetry does not play any role, and so for the moment we will leave the
group $G$ completely generic, thereby capturing the most general situation in
$D=3$.

The original formulation \cite{Nicolai:2000sc,Nicolai:2001sv} of
maximal gauged $D=3$ supergravity requires the introduction of gauge
vectors $A_{\mu}{}^{M}$ transforming in the adjoint representation
of $G$ which do not describe new degrees of freedom but are dual to
scalars.  Owing to this fact, the embedding tensor carries in three
dimensions two adjoint indices and thus reads $\Theta_{MN}$. More
precisely, the gauge vectors enter via a topological Chern-Simons
term, whose invariant tensor is precisely given by $\Theta_{MN}$
(cf.~(\ref{defaction}) below). In this case, the embedding tensor is
symmetric, $\Theta_{MN}=\Theta_{NM}$, and the tensors defined in
(\ref{Xtensor}) read

\begin{equation}
X_{ M N}{}^{ P} \ = \ \Theta_{ M K}f^{ K P}{}_{ N} \ = \ X_{[ M
N]}{}^{ P}+Z^{ P}{}_{ M N}\;, \qquad Z^{ P}{}_{ M N} \ = \ \Theta_{
K( M}f^{ K P}{}_{ N)}\;,
\end{equation}

\noindent with the structure constants of $G$ satisfying $[t^{
M},t^{ N}]=-f^{ M N}{}_{ K}t^{ K}$. As in (\ref{quadconstr2}), the
quadratic constraint states that the symmetric part $Z$ vanishes
upon contraction with the embedding tensor, $\Theta_{ P K}Z^{ P}{}_{
M N}=0$. Ultimately, supersymmetry requires in addition a linear
constraint. However, for the bosonic gauge covariance of the tensor
hierarchy this constraint is immaterial and thus it is sufficient to
impose only the quadratic constraints.

For the present purpose it suffices to inspect the equations of motion of the
gauge vectors. By virtue of the Chern-Simons term they take the form of
first-order duality relations,

\begin{equation}
\label{contrdual} e^{-1}\varepsilon^{\mu\nu\rho}\Theta_{ M
N}F_{\nu\rho}{}^{ N} \ = \
-2\Theta_{ M N}J^{\mu N}\;.
 \end{equation}

\noindent
Here, the current $J^{\mu M}$ corresponds to the Noether
current of the ungauged theory, which can be written in terms of the
Killing vector fields $k_{i}{}^{ M}(\phi)$ generating $G$ as

\begin{equation}\label{KillingDef}
J_{\mu}{}^{ M} \ = \ D_{\mu}\phi^{i}\;k_{i}{}^{ M}\;,
\end{equation}

\noindent
where $i,j,\ldots$ are the coordinate labels of the scalar
manifold. The field strength takes the standard form

\begin{eqnarray}
\label{fieldstrength} F_{\mu\nu}{}^{ M} \ = \
\partial_{\mu}A_{\nu}{}^{ M}-\partial_{\nu}A_{\mu}{}^{ M}
+X_{ N P}{}^{ M}A_{[\mu}{}^{ N}A_{\nu]}{}^{ P}\;,
\end{eqnarray}

\noindent
where the quadratic term has to be antisymmetrized explicitly due to the lack
of antisymmetry of the `structure constant' $X_{ N P}{}^{ M}$.

At this stage the situation is very similar to the four-dimensional case discussed in the
previous section. Due to the simplicity in $D=3$, it is instructive to repeat below a few of the remarks
we already made in the previous section. First, one may wonder whether it is possible  to obtain the
scalar equations of motion from the duality relation \eqref{contrdual} by acting on it with a
derivative $D_{\mu}$. This turns out not to be the case, since
(\ref{contrdual}) is only a projected version of the naive duality relation in
that both sides appear contracted with the embedding tensor. In fact, in
gauged supergravity there is a scalar potential, whose contributions to the
scalar field equations are invisible upon contraction with the embedding
tensor. Thus, the duality relation obtained from the action does not imply the
scalar field equations, though it is nevertheless compatible with them. One
might be tempted to impose the unprojected duality relations by dropping the
contraction with $\Theta_{ M N}$, in order to obtain the full field
equations. However, there are two immediate obstacles. First, the naive
Bianchi identity $D_{[\mu}F_{\nu\rho]}{}^{ M}=0$ required for deriving
second-order equations as integrability conditions holds for the field
strength in (\ref{fieldstrength}) only upon contraction with $\Theta_{ M N}$.
Second, it is clear that the contributions from a scalar potential cannot be
reproduced in this way, due to the fact that one cannot `pull out a
derivative' of the scalar potential. It turns out that the resolution of these
two problems is related and naturally suggested by the structure of the tensor
hierarchy. Specifically, this will introduce higher-rank tensor fields that
allow for covariant field strengths satisfying consistent Bianchi identities.
Moreover, these additional tensor fields will be accompanied by further
duality relations which encode, in particular, the scalar potential. This set
of first-order field equations defines the duality hierarchy which will be
discussed in the next subsection.


\subsection{The $D=3$ tensor and duality hierarchy}
\label{duality hierarchy}

As in the $D=4$ case, the tensor hierarchy can be systematically introduced by
requiring that the field strengths satisfy Bianchi identities and transform
covariantly according to their index structure. First, we modify the field
strength (\ref{fieldstrength}) by a St\"uckelberg-like coupling involving a
2-form gauge potential $B^{NK} = B^{KN}$,

\begin{equation}
\label{modstrength} H_{\mu\nu}{}^{ M} \ = \ F_{\mu\nu}{}^{ M}-2Z^{
M}{}_{ N K} B_{\mu\nu}{}^{ N K}\;.
\end{equation}

\noindent
By virtue of the quadratic constraint (\ref{quadconstr2}) the extra term
vanishes upon contraction with the embedding tensor.  Thus, all non-covariant
terms in the variation of the (unprojected) $F_{\mu\nu}{}^{ M}$ can be
absorbed into a suitable variation of the 2-form potential. Specifically,
under the standard form of the gauge transformation

\begin{equation}
\label{Avar} \delta A_{\mu}{}^{ M} \ = \ D_{\mu}\Lambda^{ M} \ = \
\partial_{\mu}\Lambda^{ M}
+X_{ N P}{}^{ M}A_{\mu}{}^{ N}\Lambda^{ P}\, ,
\end{equation}

\noindent
one finds

\begin{equation}
\delta F_{\mu\nu}{}^{ M} \ = \ X_{ N P}{}^{ M}F_{\mu\nu}{}^{
N}\Lambda^{ P} -2Z^{ M}{}_{ N P}A_{[\mu}{}^{ N} \delta A_{\nu]}{}^{
P}\;.
\end{equation}

We note that upon contraction with $\Theta_{ M N}$ the second term vanishes
and the structure constant in the first term is antisymmetric. In particular
the latter property is required in order to derive the standard covariant form
of the gauge transformation. The lack of covariance for the unprojected field
strength can now be compensated by assigning gauge transformations to the
2-form in (\ref{modstrength}). Requiring the covariant variation

\begin{equation}
\delta H_{\mu\nu}{}^{ M} \ = \ -\Lambda^{ N}X_{ N P}{}^{
M}H_{\mu\nu}{}^{ P}\;,
\end{equation}

\noindent
determines the 2-form gauge variation, with parameter $\Lambda^{MN} = \Lambda^{NM}$, to be

\begin{equation}
\label{2formvar} \delta B_{\mu\nu}{}^{ M N} \ = \
D_{[\mu}\Lambda_{\nu]}{}^{ M N} -A_{[\mu}{}^{\langle M}\delta
A_{\nu]}{}^{ N\rangle} +\Lambda^{\langle M}H_{\mu\nu}{}^{
N\rangle}+\cdots\;,
 \end{equation}

\noindent
up to terms that vanish upon contraction with $Z^{ P}{}_{ M N}$.  Here, the
brackets $\langle\;\rangle$ a priori denote ordinary (unit-strength)
symmetrization. However, Eq.~(\ref{2formvar}) and all subsequent relations
directly generalize to the case, where a linear constraint has been imposed,
for which $\langle\;\rangle$ has to be interpreted as the corresponding
projector onto the surviving representations. We have also added the variation
of the 2-form under its own gauge parameter $\Lambda_{\mu}{}^{ M
  N}$. Invariance of (\ref{modstrength}) then requires that $A_{\mu}{}^{ M}$
transforms (as a shift) under this symmetry, i.e., the gauge variation
(\ref{Avar}) has to be modified by $\delta^{\prime}A_{\mu}{}^{ M}=Z^{ M}{}_{ N
  P}\Lambda_{\mu}{}^{ N P}$.

In a next step one can introduce a 3-form field strength $G_{\mu\nu\rho}{}^{ M
  N}$ for the 2-form gauge potential by requiring gauge covariance. It turns
out, however, to be more convenient to determine the leading terms of this
field strength by requiring that the modified field strength for the original
gauge vector satisfies a Bianchi identity,

\begin{eqnarray}
\label{Bianchi} D_{[\mu}H_{\nu\rho]}{}^{ M} \ = \ -2Z^{ M}{}_{ N
P}G_{\mu\nu\rho}{}^{ N P}\;.
\end{eqnarray}

\noindent
This uniquely determines the field strength up to terms that vanish by
contraction with $Z^{ P}{}_{ M N}$. Ultimately, we want to write covariant
duality relations involving the uncontracted $G_{\mu\nu\rho}{}^{ M N}$. As
before, this can be achieved via introducing a new potential, which is a
3-form, and assigning appropriate gauge transformations to it. Without
repeating the detailed steps of the derivation, we simply state the
results. (For more details we refer the reader to \cite{deWit:2008ta}.) The
3-form field strength reads

\begin{eqnarray}
\label{Gstrength} G_{\mu\nu\rho}{}^{ M N} & = &
D_{[\mu}B_{\nu\rho]}{}^{ M N} -A_{[\mu}{}^{\langle
M}\partial_{\nu}A_{\rho]}{}^{ N\rangle} -\frac{1}{3}X_{ K
L}{}^{\langle M} A_{[\mu}{}^{ N\rangle}A_{\nu}{}^{ K}A_{\rho]}{}^{
L}
\nonumber \\
& & \nonumber \\
& & -\frac{2}{3}Y^{ M N}{}_{ P, K L}\; C_{\mu\nu\rho}{}^{ P, K L}\;.
\end{eqnarray}

\noindent
Here, we have introduced the intertwining $Y$-tensor \cite{deWit:2008ta}

\begin{equation}
\label{intertwiner} Y^{ M N}{}_{ P, K L} \ = \ Z^{\langle M}{}_{ K
L}\delta^{ N\rangle}{}_{ P} -X_{ P\langle K}{}^{\langle M} \delta^{
N\rangle}{}_{ L\rangle}\;,
\end{equation}

\noindent
which relates the irreducible representation in which $B_{\mu\nu}{}^{ M N}$
transforms to the irreducible representation of the 3-form.

Summarizing, we find that the 2-from field strengths (\ref{modstrength}) and
the 3-form field strengths (\ref{Gstrength}) transform covariantly under the
following gauge transformations of the $D=3$ tensor hierarchy:

\begin{eqnarray}
\delta A_{\mu}{}^{ M} & = & D_{\mu}\Lambda^{ M} +Z^{ M}{}_{ N
P}\Lambda_{\mu}{}^{ N P}
\\
& & \nonumber \\
\delta B_{\mu\nu}{}^{ M N} & = & D_{[\mu}\Lambda_{\nu]}{}^{ M N}-
A_{[\mu}{}^{\langle M}\delta A_{\nu]}{}^{ N\rangle}+\Lambda^{\langle
M}H_{\mu\nu}{}^{ N\rangle} +\frac{2}{3}Y^{ M N}{}_{ P, K L}
\Lambda_{\mu\nu}{}^{ P, K L}
\\
& & \nonumber \\
\delta C_{\mu\nu\rho}{}^{ P, M N} & = &
D_{[\mu}\Lambda_{\nu\rho]}{}^{ P, M N}-3\;\delta A_{[\mu}{}^{\langle
P}B_{\nu\rho]}{}^{ M N\rangle} +A_{[\mu}{}^{\langle P}A_{\nu}{}^{ M}
\delta A_{\rho]}{}^{ N\rangle}
\nonumber \\
& & \nonumber \\
&& -\frac{3}{2}H_{[\mu\nu}{}^{\langle P}\Lambda_{\rho]}{}^{ M
N\rangle} -3\Lambda^{\langle P} G_{\mu\nu\rho}{}^{ M N\rangle}\;.
\end{eqnarray}

\noindent
Again, the brackets $\langle\;\rangle$ generically impose the constraints on
the 2-form and, via (\ref{intertwiner}), also the corresponding constraints on
the 3-form. As in $D=4$ the above gauge transformations of the $D=3$ tensor
hierarchy close off-shell.  In three dimensions the tensor hierarchy
terminates at this point, as there are no higher-rank tensor fields and no
further non-trivial Bianchi identities beyond the 3-form identity
(\ref{Bianchi}).

Now we are in a position to impose manifestly gauge-covariant duality
relations, whose compatibility conditions with the Bianchi identities
reproduce the supergravity equations of motion (up to the Einstein
equation). First, we introduce the unprojected form of the duality relation
(\ref{contrdual}), in which the field strength gets modified according to
(\ref{modstrength}),

\begin{equation}
\label{duality1} \mathcal{E}^{\mu\; M} \ \equiv \
e^{-1}\varepsilon^{\mu\nu\rho}H_{\nu\rho}{}^{ M} +2 J^{\mu M} \ = \
0\;.
\end{equation}

\noindent
Next, we define a duality relation for the 2-form potential, which introduces
the derivative of the scalar potential with respect to $\Theta$,

\begin{equation}
\label{duality2} \mathcal{E}^{ M N} \ \equiv \
e^{-1}\varepsilon^{\mu\nu\rho}G_{\mu\nu\rho}{}^{ M N}
+\frac{1}{4}G^{ M N, K L}\Theta_{ K L} \ = \ 0\;.
\end{equation}

\noindent
Here, $G^{ M N, K L}$ is a (scalar-dependent) matrix fixed by supersymmetry
(for the explicit form in case of ${\cal N}=16$ see \cite{Bergshoeff:2008qd}),
which determines the potential according to

\begin{equation}
V \ = \ \frac{1}{32}G^{ M N, K L} \Theta_{ M N}\Theta_{ K L}\;.
\end{equation}

\noindent
The (formal) G-invariance implies the following identity

\begin{equation}
k^{i M}\frac{\partial V}{\partial \phi^{i}} -2Z^{ M}{}_{ N
P}\frac{\partial
 V}{\partial \Theta_{ N P}} \ = \ 0\;.
\end{equation}

The claim is that the $D=3$ duality hierarchy (\ref{duality1}) and
(\ref{duality2}) encodes the equations of motion up to the Einstein
equations. In this example there are just two equations of motion: the vector
equations (\ref{contrdual}), resulting from (\ref{duality1}) by contracting
with $\Theta_{ M N}$, and the scalar equations of motion,

\begin{equation}
\label{scalareom0}
 D_{\mu}\left(g_{ij} D^{\mu}\phi^{j}\right) \ = \
 -2\;\frac{\partial V}{\partial\phi^{i}}\;,
\end{equation}

\noindent
where $g_{ij}$ is the metric on the scalar manifold. By acting with $D_{\mu}$
on (\ref{duality1}) and using the second duality relation (\ref{duality2}) one
obtains as a consequence of the Bianchi identity (\ref{Bianchi})

\begin{equation}\label{scalareom}
D_{\mu}J^{\mu M} \ = \ -2k^{i M}\frac{\partial
V}{\partial\phi^{i}}\;.
\end{equation}

\noindent
Alternatively, these equations are identical to the 3-form Bianchi identity
(\ref{Bianchi}) after replacing in (\ref{scalareom}) the scalar-dependent
Noether current $J^{\mu M}$ by the scalar-independent 2-form field strength
$H^{ M}$ via the duality relation (\ref{duality1}) and after replacing the
scalar-dependent (derivative of) the scalar potential $V$ by the
scalar-independent 3-form field strength $G^{ M N}$ via the duality relation
(\ref{duality2}). These second-order `conservation equations' can be viewed as
projected scalar equations of motion in the sense that (\ref{scalareom})
results from (\ref{scalareom0}) by contracting with the Killing vector
$k_{i}{}^{M}$.


\subsection{The $D=3$ action}
\label{action}

An action including all fields of the $D=3$ tensor hierarchy has
already been constructed in \cite{Bergshoeff:2008qd,deWit:2008ta}
(see \cite{Bergshoeff:2008cz} for the case of global supersymmetry).
It reads

\begin{equation}
\label{defaction}
\begin{split}
{\cal L} \ = \ & -\frac{1}{4} eR + \frac{1}{4} e P^{\mu a}P_{\mu a}
- eV
\\
& -\frac{1}{4}\varepsilon^{\mu\nu\rho}A_{\mu}{}^{ M}\vartheta_{ M N}
\left(\partial_{\nu}A_{\rho}{}^{ N} +\frac{1}{3} X_{ K L}{}^{
N}A_{\nu}{}^{ K} A_{\rho}{}^{ L}\right)+{\cal L}_{\rm fermions}
\\
& +\frac{1}{4} \varepsilon^{\mu\nu\rho}D_{\mu}\vartheta_{ M N}
B_{\nu\rho}{}^{ M N} +\frac{1}{6}\vartheta_{ P K}Z^{ K}{}_{ M N}
\varepsilon^{\mu\nu\rho}C_{\mu\nu\rho}{}^{ P, M N}\;.
\end{split}
\end{equation}

\noindent
Here we used the definition

\begin{equation}\label{P}
P_{\mu}{}^{a} \ = \ D_{\mu}\phi^{i}\;e_{i}{}^{a}(\phi)\;,
\end{equation}

\noindent
where $e_{i}{}^{a}$ denotes the vielbein on the scalar manifold with flat
indices $a,b,\ldots$.  We denote the embedding tensor by $\vartheta_{ M
  N}=\vartheta_{ M N}(x)$ in order to indicate that it is now a space-time
dependent field. As long as the precise form of the scalar potential and the
fermionic couplings is not specified, this form of the action is completely
general and applies to all gauged supergravities in $D=3$. In particular, the
scalar kinetic term represents a generic non-linear sigma model.

In (\ref{defaction}) we have made use of the fact that the 2-form potentials
emerging in the tensor hierarchy carry the same $G$-representation as the
embedding tensor. This follows from the fact that the tensor $Z^{ M}{}_{ N K}$
contracting the 2-forms in (\ref{modstrength}) can be viewed as a $G$-rotation
of $\vartheta_{ M N}$ and thus satisfies the same representation constraint
(if any) as the embedding tensor. The space-time dependent embedding tensor
$\vartheta_{ M N}(x)$ in (\ref{defaction}) is set to a constant satisfying the
quadratic constraints by the field equations for the 2- and 3-forms.

Like in $D=4$, in principle, it is also possible to enforce linear constraints
via additional top-form Lagrange multipliers. However, since for the action in
$D=3$ the linear constraint is immaterial for bosonic gauge invariance, this
would be redundant and so we will not follow this route here. This is in
contrast to the $D=4$ case where linear constraints do play a role for bosonic
gauge invariance. In that case we did introduce a Lagrange multiplier for the
linear constraint.

In this reformulation with dynamical embedding tensor the original invariance
of the action is violated by terms proportional to $\partial_{\mu}\vartheta_{
  M N}$ and the quadratic constraint. This can be compensated by assigning
appropriate gauge transformations to the 2- and 3-form, as has been done in
\cite{Bergshoeff:2008qd}. The corresponding gauge variations will be denoted
by ${\delta}_a$ in order to distinguish them from the gauge transformations
$\delta_h$ of the tensor hierarchy. As in $D=4$ we find that $\delta_a$ and
$\delta_h$ differ by terms that are proportional to the duality relations
(\ref{duality1}) and (\ref{duality2}):

 \begin{eqnarray}
\label{modgauge} {\delta}_aB_{\mu\nu}{}^{ M N} & = & \delta_h
B_{\mu\nu}{}^{ M N}+\frac{1}{2}e
\varepsilon_{\mu\nu\rho}\Lambda^{\langle M} \mathcal{E}^{\rho\;
N\rangle}\;,
\nonumber \\
& & \nonumber \\
{\delta}_aC_{\mu\nu\rho}{}^{ P, M N} & = & \delta_h
C_{\mu\nu\rho}{}^{ P, M N}
-\frac{3}{4}e\varepsilon_{\sigma[\mu\nu}\mathcal{E}^{\sigma\langle
P} \Lambda_{\rho]}{}^{ M N\rangle}
-\frac{1}{2}e\varepsilon_{\mu\nu\rho}\Lambda^{\langle
P}\mathcal{E}^{ M N\rangle}\;,
\end{eqnarray}

\noindent
as can be inferred from \cite{Bergshoeff:2008qd} by comparing the ${\delta}_a$
variations with the tensor hierarchy.\footnote{Strictly speaking, only the
  maximally supersymmetric case has been investigated in
  \cite{Bergshoeff:2008qd}. However, as far as invariance of the bosonic
  Lagrangian is concerned, this is no restriction.} We note that the
variations of the original vectors and scalars remain unchanged. This
modification is precisely such that all field strengths in the transformation
rules get replaced by dual (matter) contributions, as Noether currents, etc.

Let us stress again that (\ref{modgauge}) is not equivalent to the original
gauge transformations of the tensor hierarchy. First of all, (\ref{modgauge})
does not represent a modification by an equations-of-motion symmetry, since
this would have to act on all fields and not just the 2- and
3-forms. Moreover, the modified gauge transformations are not even on-shell
equivalent to the tensor hierarchy, due to the fact that neither the duality
relation (\ref{duality1}) nor (\ref{duality2}) follows from the action. More
precisely, the field equations are

\begin{eqnarray}
  \frac{\delta S}{\delta A_{\mu}{}^{M}} &=&
  -\frac{1}{4}\vartheta_{MN}\;{\cal E}^{\mu\; N} \ = \ 0\;, \\
  \label{thetaeq}
  \frac{\delta S}{\delta \vartheta_{MN}} &=&
  -\frac{1}{4}\left({\cal E}^{MN}+A_{\mu}{}^{\langle M}{\cal
  E}^{\mu\; N\rangle}\right) \ = \ 0\;.
\end{eqnarray}

\noindent Thus, the first duality relation appears only in a
contracted version. Once its unprojected form (\ref{duality1}) has
been imposed by hand, the field equations for the embedding tensor
(\ref{thetaeq}) turn out to be equivalent to (\ref{duality2}). As a
consequence, the field equations obtained from the action are not
manifestly gauge-covariant but rather rotate under the gauge
transformations in a highly intricate way into the other field
equations (including second-order matter equations)
\cite{Bergshoeff:2008qd}. Moreover, the off-shell closure of the
gauge algebra characteristic for the abstract tensor hierarchy is
violated in that closure requires the validity of \textit{all} field
equations (except the Einstein equation).


\section{Conclusions}
\label{sec-conclusions}

In this paper we have showed how the second-order $p$-form equations of motion
and the projected scalar equations of motion of general $D=3,4$ gauged
supergravity theories\footnote{Actually, our results should apply, unmodified,
  to more general $D=3,4$ theories with no supersymmetry.} can be derived by a
\textit{duality hierarchy}, i.e.~a set of first-order duality relations
between $p$-form curvatures.

Our starting point has been the complete \textit{tensor hierarchy} of the
embedding tensor formalism which we have used to derive the off-shell gauge
algebra for a set of $p$-form potentials, not including the scalars and the
metric tensor. Next, in a second step we have put the tensor hierarchy
on-shell by introducing duality relations between the curvatures of the tensor
hierarchy.  These duality relations contain the metric tensor and all the
information about the scalar couplings via natural objects, like the Noether
current, the derivative of the scalar potential with respect to the embedding
tensor and, in the case of four dimensions, a function describing the
scalar-vector couplings.  We have showed how the duality relations, together
with the Bianchi identities of the tensor hierarchy, lead to the desired
second-order equations of motion for the $p$-form potentials and to the
projected equations of motion for the scalars.

In a third and final step we have constructed a gauge-invariant
action for all the fields of the tensor hierarchy. Here a subtlety
occurred. We find that the gauge transformations of the action, with
on-shell closed gauge algebra, are not the same as the gauge
transformations of the tensor hierarchy, with off-shell closed gauge
algebra. They differ by (unprojected) duality relations some of
which do not follow from extremizing the action although they are
part of the duality hierarchy. We find that the transformation rules
that leave the action invariant are obtained from the transformation
rules of the tensor hierarchy by replacing everywhere curvatures by
dual curvatures via the duality relations except in one term in the
gauge transformations of the 4-forms $D^{NPQ}$, associated to the
linear constraint. This exception to the almost-general rule
disappears if one solves the linear constraint at the beginning and
uses only the allowed field representations. It is reasonable to
conjecture that the same will be true in other dimensions and, if
true, it would be interesting to find an explanation for this
general pattern. It would also be interesting to find out how the
general $D=4$ tensor hierarchy is modified if one relaxes the linear
constraint as in Ref.~\cite{DeRydt:2008hw}, in which the classical
lack of gauge invariance can be compensated by a quantum anomaly.

It is natural to ask under which circumstances the duality hierarchy can give
rise to the {\sl full} set of scalar equations of motion. For this to be the
case, the Killing vector fields need to be `left-invertible'. For instance, in
the $D=3$ example this means that (\ref{scalareom}) implies
(\ref{scalareom0}). A necessary condition is that the dimension of the
isometry group is larger or equal to the dimension of the scalar
manifold. This is satisfied for coset manifolds $G/H$. In order to see this,
let ${\cal V}$ be $G$-valued and $P_{\mu}{}^{a}=[{\cal V}^{-1}D_{\mu}{\cal
  V}]^{a}$ the coset part of the $G$-invariant Maurer-Cartan forms.  The
Noether current results from $P_{\mu}{}^{a}$ by converting the flat index to a
curved or rigid one by means of the coset vielbein ${\cal V}$,

\begin{equation}
  J_{\mu}{}^{M} \ = \ {\cal V}^{M}{}_{a}P_{\mu}{}^{a}\;,
\end{equation}

\noindent
where the contraction is only over the `coset directions'.
Comparing with (\ref{KillingDef}) one infers

 \begin{equation}
  k_{i}{}^{M} \ = \ {\cal V}^{M}{}_{a}\; e_{i}{}^{a}\;.
 \end{equation}

\noindent Since the vielbeine $e$ and ${\cal V}$ are both invertible
the desired result follows. Thus, in case of supergravity theories
based on coset manifolds, the entire set of field equations (except
the Einstein equations) are encoded in first-order duality
relations.

It is tempting to conjecture that this pattern will persist in general
dimensions $D>4$. In the context of higher dimensions it is noteworthy that to
construct an action not always all fields of the tensor hierarchy are
involved. Apart from low-rank forms, which are required for consistent
gaugings, and the $(D-1)$- and $D$-forms, which can be interpreted as Lagrange
multipliers, there appears a gap `in between'. For instance, the $D=5$ gauged
supergravity actions of \cite{deWit:2002vt,deWit:2004nw} do not contain a
3-form.  In contrast, at the level of the duality hierarchy one is forced to
introduce this 3-form in order to recover the correct second-order field
equations \cite{Riccioni:2007ni}.

One may wonder whether it is possible to also obtain the Einstein equations as
compatibility conditions from duality relations. Remarkably, this turns out to
be possible upon introducing the dual graviton transforming in the mixed-Young
tableaux representation $(D-3,1)$, as has been shown recently
\cite{Boulanger:2008nd}. At first sight one would think that one cannot write
first-order duality relations since it is not possible to `pull out a
derivative' of the energy-momentum tensor \cite{Bergshoeff:2008vc}. This is
similar to the scalar equations of motion discussed in this paper, where it
was not possible to pull out a derivative of the scalar potential.  The
resolution to this obstruction is in precise analogy to the scalar equations:
it requires the introduction of an extra higher rank tensor field, which in
this case contains the $(D-2,1)$ Young tableaux. Thus, like in
(\ref{duality2}), a second duality relation has to be imposed, that explicitly
contains the energy-momentum tensor. It is intriguing that, therefore,
\textit{all} supergravity equations can be written as first-order duality
relations (assuming a sufficiently large symmetry in the scalar sector).

Finally, it is interesting to contemplate the possible relation of our
findings to the $E_{11}$ approach to supergravity
\cite{West:2001as,Schnakenburg:2001ya,Kleinschmidt:2003mf,Riccioni:2007ni}. In
that context the formulation in terms of duality relations seems to be more
natural and thus the present analysis may be of relevance. In this context we
note the different status of the higher $p$-forms in the action and the
duality hierarchy. For instance, the incorporation of the top-form and
next-to-top form potentials in an action leads to complicated gauge
transformation rules with an on-shell closed gauge algebra
\cite{Bergshoeff:2008qd}. It is unlikely that such a structure has a direct
Kac-Moody origin.  In contrast, the gauge symmetries realized on the duality
relations close off-shell in agreement with the tensor hierarchy, and
therefore a possible connection to Kac-Moody algebras appears to be more
promising.  The Kac-Moody approach to supergravity has only been developed so
far for supergravities whose scalar sector is given by a coset manifold. It is
precisely for these cases that the duality hierarchy reproduces the full set
of scalar equations of motion and not just the projected ones. It would be of
interest to extend both the Kac-Moody approach as well as the duality
hierarchy to supergravities whose scalar sector is given by more general
manifolds.

\textit{Note added:} We would like to mention ref.~\cite{BdeWit},
which was brought to our attention after this paper has been
submitted to the bulletin board. Section 4 of \cite{BdeWit} also
deals with the $D=4$ tensor hierarchy and has some overlap with our
section 3.


\section*{Acknowledgments}

MH would like to thank the \'Ecole Normale Superieure of Lyon for
its hospitality during the early stages of this work and
H.~Samtleben for many useful discussions during her stay there. TO
would like to thank the Center for Theoretical Physics of the
University of Groningen for its hospitality. JH was supported by a
research grant of the Swiss National Science Foundation and wishes
to thank the Instituto de F\'{i}sica Te\'{o}rica of the Universidad
Aut\'{o}noma de Madrid for its hospitality. The Center for Research
and Education in Fundamental Physics is supported by the
``Innovations- und Kooperationsprojekt C-13'' of the Schweizerische
Universitaetskonferenz SUK/CRUS. This work has been supported in
part by the INTAS Project 1000008-7928, the Spanish Ministry of
Science and Education grants FPU AP2004-2574 (MH), FPA2006-00783 (MH
and TO), the Comunidad de Madrid grant HEPHACOS P-ESP-00346 (MH and
TO), the Spanish Consolider-Ingenio 2010 program CPAN CSD2007-00042
(MH and TO) and by the EU Research Training Network
\textit{Constituents, Fundamental Forces and Symmetries of the
  Universe} MRTN-CT-2004-005104.
This work is part of the research programme of the `Stichting voor
Fundamenteel Onderzoek der Materie (FOM)'.
 Further, TO wishes to express his gratitude
to M.M.~Fern\'andez for her unwavering support.

\appendix

\section{Properties of the $W$ tensors}
\label{app-wtensorproperties}

The $W$ tensors defined in Eqs.~(\ref{eq:W1})-(\ref{eq:W3}) satisfy
the following properties:

\begin{eqnarray}
\label{eq:zW1}
\Theta_{M}{}^{C} W_{C}{}^{MAB} & = & 2Q^{AB}\, ,\\
& & \nonumber \\
\label{eq:zW2}
\Theta_{M}{}^{C} W_{CNPQ}{}^{M} & = & L_{NPQ}\, ,\\
& & \nonumber \\
\label{eq:zW3} \Theta_{M}{}^{C}W_{CNP}{}^{EM} & = &
2Q_{NP}{}^{E}\, ,
\end{eqnarray}

\begin{eqnarray}
\label{eq:dQ1} \frac{\partial Q^{AB}}{\partial \Theta_{M}{}^{C}}
& = &
W_{C}{}^{MAB} \, ,\\
& & \nonumber \\
\label{eq:dQ2} \frac{\partial L_{NPQ}}{\partial \Theta_{M}{}^{C}}
 & = &
W_{CNPQ}{}^{M}\, ,\\
& & \nonumber \\
\label{eq:dQ3} \frac{\partial Q_{NP}{}^{E}}{\partial
\Theta_{M}{}^{C}}
 & = &
W_{CNP}{}^{EM} \, ,
\end{eqnarray}

\begin{eqnarray}
\label{eq:dzW1} \delta \Theta_{M}{}^{C} W_{C}{}^{MAB} & = &
\Theta_{M}{}^{C} \delta W_{C}{}^{MAB}  =
{\textstyle\frac{1}{2}}\delta (\Theta_{M}{}^{C}
W_{C}{}^{MAB})=\delta Q^{AB}\, ,\\
& & \nonumber \\
\label{eq:dzW2}
\delta \Theta_{M}{}^{C} W_{CNPQ}{}^{M} & = & \delta L_{NPQ}\, , \\
& & \nonumber \\
\label{eq:dzW3} \delta \Theta_{M}{}^{C} W_{CNP}{}^{EM} & = &
\Theta_{M}{}^{C} \delta  W_{CNP}{}^{EM} ={\textstyle\frac{1}{2}}
\delta (\Theta_{M}{}^{C} W_{CNP}{}^{EM}) =\delta Q_{NP}{}^{E}\, ,
\end{eqnarray}

\noindent where $Q^{AB}$, $Q_{NP}{}^{E}$ and $L_{NPQ}$ are the
quadratic and linear constraints Eqs.~(\ref{eq:quadraticE}),
(\ref{eq:quadraticTdef}) and (\ref{eq:linear}) imposed on the
embedding tensor and where we have not used the constraints
themselves.


\section{Transformations and field strengths in the $D=4$ tensor hierarchy}
\label{app-gaugetranshierarchy}

The gauge transformations of the different fields of the tensor
hierarchy are

\begin{eqnarray}
\delta_{h} A^{M} & = & -\mathfrak{D}\Lambda^{M} -Z^{MA}\Lambda_{A}\,
,
\\
& & \nonumber \\
\delta_{h} B_{A} & = &  \mathfrak{D}\Lambda_{A} +2T_{A\,
NP}[\Lambda^{N}F^{P} +{\textstyle\frac{1}{2}}A^{N}\wedge\delta_{h}
A^{P}] -Y_{AM}{}^{C}\Lambda_{C}{}^{M}\, ,
\\
& & \nonumber \\
\delta_{h} C_{A}{}^{M} & = & \mathfrak{D}\Lambda_{A}{}^{M}
 -F^{M}\wedge\Lambda_{A}
 -\delta_{h} A^{M}\wedge B_{A}
-{\textstyle\frac{1}{3}}T_{A\, NP} A^{M}\wedge A^{N} \wedge
\delta_{h} A^{P} +\Lambda^{M}H_{A}
\nonumber \\
& & \nonumber \\
& & -W_{A}{}^{MAB}\Lambda_{AB} -W_{ANPQ}{}^{M}\Lambda^{NPQ}
-W_{ANP}{}^{EM}\Lambda_{E}{}^{NP}\, ,
\nonumber \\
& & \nonumber \\
\delta_{h} D_{AB} & = & \mathfrak{D}\Lambda_{AB} +\alpha
B_{[A}\wedge Y_{B]P}{}^{E}\Lambda_{E}{}^{P}
+\mathfrak{D}\Lambda_{[A}\wedge B_{B]} -2\Lambda_{[A}\wedge H_{B]}
\nonumber \\
& & \nonumber \\
& & +2T_{[A| NP}[\Lambda^{N}F^{P}
-{\textstyle\frac{1}{2}}A^{N}\wedge \delta_{h} A^{P}]\wedge B_{|B]}\, , \\
& & \nonumber \\
\delta_{h} D_{E}{}^{NP} & = & \mathfrak{D}\Lambda_{E}{}^{NP}
-[F^{N}-{\textstyle\frac{1}{2}}(1-\alpha)Z^{NA}B_{A}]\wedge
\Lambda_{E}{}^{P}
\nonumber \\
& & \nonumber \\
& & +C_{E}{}^{P}\wedge \delta_{h} A^{N} +{\textstyle\frac{1}{12}}
T_{EQR} A^{N}\wedge A^{P} \wedge A^{Q} \wedge \delta_{h} A^{R}
+\Lambda^{N}G_{E}{}^{P}\, ,\\
& & \nonumber \\
\delta_{h} D^{NPQ} & = & \mathfrak{D}\Lambda^{NPQ} -2A^{(N}\wedge
dA^{P}\wedge \delta_{h} A^{Q)} -{\textstyle\frac{3}{4}}
X_{RS}{}^{(N} A^{P|} \wedge A^{R}\wedge A^{S} \wedge  \delta_{h}
A^{|Q)}
 \nonumber \\
& & \nonumber \\
& & -3\Lambda^{(N}F^{P}\wedge F^{Q)} \, ,
\end{eqnarray}

\noindent and their gauge-covariant field strengths are

\begin{eqnarray}
F^{M} & = & dA^{M} +{\textstyle\frac{1}{2}}X_{[NP]}{}^{M}A^{N}\wedge
A^{P} +Z^{MA}B_{A}\, ,
\\
& & \nonumber \\
H_{A} & = & \mathfrak{D}B_{A} +T_{A\, RS}A^{R}\wedge[dA^{S}
+{\textstyle\frac{1}{3}}X_{NP}{}^{S}A^{N}\wedge A^{P}]
+Y_{AM}{}^{C}C_{C}{}^{M}\, , \label{eq:HA}
\\
& & \nonumber \\
G_{C}{}^{M} & = & \mathfrak{D}C_{C}{}^{M}
+[F^{M}-{\textstyle\frac{1}{2}}Z^{MA}B_{A}]\wedge B_{C}
+{\textstyle\frac{1}{3}}T_{C\, SQ} A^{M}\wedge A^{S} \wedge dA^{Q}
\nonumber\\
& & \nonumber \\
& & +{\textstyle\frac{1}{12}}T_{C\, SQ}X_{NT}{}^{Q}A^{M}\wedge A^{S}
\wedge A^{N} \wedge A^{T}
\nonumber \\
& & \nonumber \\
& & +W_{C}{}^{MAB}D_{AB} +W_{CNPQ}{}^{M}D^{NPQ} +W_{CNP}{}^{EM}
D_{E}{}^{NP}\, . \label{eq:GCM}
\end{eqnarray}

These field strengths are related by the following hierarchical
Bianchi identities

\begin{eqnarray}
\mathfrak{D}F^{M} & = & Z^{MA}H_{A}\, ,
\\
& & \nonumber \\
\mathfrak{D}H_{A} & = & Y_{AM}{}^{C} G_{C}{}^{M}  +T_{A\,
MN}F^{M}\wedge F^{N}\, .
\end{eqnarray}


\section{Gauge transformations in the $D=4$ duality hierarchy and action}
\label{app-gaugetranacion}

In hierarchy variables, the total action takes the form

\begin{equation}
\label{eq:totalaction}
\begin{array}{rcl}
S  & = &  {\displaystyle\int} \left\{ \star R
-2\mathcal{G}_{ij^{*}}\mathfrak{D}Z^{i}
\wedge\star\mathfrak{D}Z^{*\, j^{*}} +2F^{\Sigma}\wedge G_{\Sigma}
-\star V \right.
\\
& & \\
& & -4Z^{\Sigma A}B_{A}\wedge \left(F_{\Sigma}
-{\textstyle\frac{1}{2}}Z_{\Sigma}{}^{B}B_{B}\right)
-{\textstyle\frac{4}{3}}  X_{[MN]\Sigma} A^{M}\wedge A^{N} \wedge
\left(F^{\Sigma} -Z^{\Sigma B}B_{B}\right)
\\
& & \\
& & -{\textstyle\frac{2}{3}}  X_{[MN]}{}^{\Sigma}A^{M}\wedge A^{N}
\wedge \left(dA_{\Sigma} -{\textstyle\frac{1}{4}} X_{[PQ]\Sigma}
A^{P}\wedge A^{Q}\right)
\\
& & \\
& & -2\mathfrak{D}\vartheta_{M}{}^{A}\wedge (C_{A}{}^{M}
+A^{M}\wedge B_{A}) +2Q_{NP}{}^{E}(D_{E}{}^{NP}
-{\textstyle\frac{1}{2}}A^{N}\wedge A^{P} \wedge B_{E})
\\
& & \\
& & \left. +2Q^{AB}D_{AB} +2L_{NPQ}D^{NPQ}
\right\}\, .\\
\end{array}
\end{equation}

A general variation of this action is given by

\begin{equation}
\begin{array}{rcl}
\delta S & = & {\displaystyle\int} \left\{ \delta g^{\mu\nu}
{\displaystyle\frac{\delta S}{\delta g^{\mu\nu}}} +\left(\delta
Z^{i} {\displaystyle\frac{\delta S}{\delta Z^{i}}}
+\mathrm{c.c.}\right) -\delta A^{M}\wedge \star
{\displaystyle\frac{\delta S}{\delta A^{M}}} +2\delta B_{A} \wedge
\star {\displaystyle\frac{\delta S}{\delta  B_{A}}} \right.
\\
& & \\
& & -2\mathfrak{D}\vartheta_{M}{}^{A}\wedge \delta C_{A}{}^{M}
+2Q_{NP}{}^{E} \delta D_{E}{}^{NP} +2Q^{AB}\delta D_{AB} +2L_{NPQ}
\delta D^{NPQ}
\\
& & \\
& & \left. +\delta \vartheta_{M}{}^{A} {\displaystyle\frac{\delta
S}{\delta  \vartheta_{M}{}^{A}}} \right\}\, ,
\end{array}
\end{equation}

\noindent where

\begin{eqnarray}
{\displaystyle\frac{\delta S}{\delta g^{\mu\nu}}} & = & \star
\mathbb{I}\left\{ G_{\mu\nu}
+2\mathcal{G}_{ij^{*}}[\mathfrak{D}_{\mu}Z^{i}
\mathfrak{D}_{\nu}Z^{*\, j^{*}} -{\textstyle\frac{1}{2}}g_{\mu\nu}
\mathfrak{D}_{\rho}Z^{i}\mathfrak{D}^{\rho}Z^{*\, j^{*}}]
-G^{M}{}_{(\mu|}{}^{\rho}\star G_{M| \nu) \rho} \right.
\nonumber \\
& & \nonumber \\
& & \left.
 +{\textstyle\frac{1}{2}}g_{\mu\nu}V
\right\}\, ,
\\
& & \nonumber \\
{\textstyle \frac{1}{2}} {\displaystyle\frac{\delta S}{\delta
Z^{i}}} & = & \mathcal{G}_{ij^{*}}\mathfrak{D}\star \mathfrak{D}
Z^{*\, j^{*}} -\partial_{i}G_{M}{}^{+}\wedge G^{M +} -\star
{\textstyle\frac{1}{2}}\partial_{i}V\, ,
\\
& & \nonumber \\
-{\textstyle\frac{1}{4}} {\displaystyle\star \frac{\delta S}{\delta
A^{M}}} & = &
\mathfrak{D}F_{M}-{\textstyle\frac{1}{4}}\vartheta_{M}{}^{A}\star
j_{A} -{\textstyle\frac{1}{3}}dX_{[PQ] M}\wedge A^{P}\wedge A^{Q}
+{\textstyle\frac{1}{2}}Q_{MP}{}^{E}C_{E}{}^{P}
-{\textstyle\frac{1}{2}}Q_{(NM)}{}^{E}A^{N}\wedge B_{E}
\nonumber \\
& & \nonumber \\
& & -L_{MNP}A^{N}\wedge \left(dA^{P}
+{\textstyle\frac{3}{8}}X_{[RS]}{}^{P}A^{R}\wedge A^{S}\right)
+{\textstyle\frac{1}{8}}Q_{NP}{}^{E}T_{E\, QM}A^{N} \wedge A^{P}
\wedge A^{Q}
\nonumber \\
& & \nonumber \\
& & -d(F_{M}-G_{M}) -X_{[MN]}{}^{P}A^{N}\wedge(F_{P}-G_{P})
+{\textstyle\frac{1}{2}}\mathfrak{D}\vartheta_{M}{}^{A}\wedge B_{A}
\, ,
\\
& & \nonumber \\
{\displaystyle\star \frac{\delta S}{\delta  B_{A}}} & = &
\vartheta^{P A}(F_{P}-G_{P}) +Q^{AB}B_{B}
-\mathfrak{D}\vartheta_{M}{}^{A}\wedge A^{M}
-{\textstyle\frac{1}{2}}Q_{NP}{}^{A}A^{N}\wedge A^{P}\, ,
\\
& & \nonumber \\
{\textstyle\frac{1}{2}} {\displaystyle \frac{\delta S}{\delta
\vartheta_{M}{}^{A}}} & = & (G_{A}{}^{M}
-{\textstyle\frac{1}{2}}\star\partial V/\partial \vartheta_{M}{}^{A}
) -A^{M}\wedge (H_{A}+{\textstyle\frac{1}{2}}\star j_{A})
\nonumber \\
& & \nonumber \\
& & +{\textstyle\frac{1}{2}}T_{ANP}A^{M}\wedge A^{N}\wedge (F^P-G^P)
-(F^M-G^M)\wedge B_{A}\, ,
\end{eqnarray}

\noindent and vanishes, up to total derivatives, for the gauge
transformations

\begin{eqnarray}
\delta_{a}\vartheta_{M}{}^{A} & = & 0\, ,
\\
& & \nonumber \\
\delta_{a}Z^{i} & = & \Lambda^{M}\vartheta_{M}{}^{A} k_{A}{}^{i}\, ,
\\
& & \nonumber \\
\delta_{a} A^{M} & = & \delta_{h} A^{M}\, ,
\\
& & \nonumber \\
\delta_{a} B_{A} & = & \delta_{h} B_{A} -2T_{A\,
NP}\Lambda^{N}(F^{P}-G^{P})\, ,
\\
& & \nonumber \\
\delta_{a} C_{A}{}^{M} & = & \delta_{h} C_{A}{}^{M}
+\Lambda_{A}\wedge  (F^M-G^M)
-\Lambda^{M}(H_{A}+{\textstyle\frac{1}{2}}\star j_{A})\, ,
\\
& & \nonumber \\
\delta_{a} D_{AB} & = & \delta_{h} D_{AB} +2\Lambda_{[A}\wedge
(H_{B]}+{\textstyle\frac{1}{2}}\star j_{B]}) -2T_{[A|\,
NP}\Lambda^{N}(F^{P}-G^{P})\wedge B_{|B]}
\, , \\
& & \nonumber \\
\delta_{a} D_{E}{}^{NP} & = & \delta_{h} D_{E}{}^{NP}
-\Lambda^{N}(G_{E}{}^{P} -{\textstyle\frac{1}{2}}\star\partial
V/\partial \vartheta_{P}{}^{E} ) +2 (F^{N}-G^{N})\wedge
\Lambda_{E}{}^{P}\, ,
\\
& & \nonumber \\
\delta_{a} D^{NPQ} & = & \delta_{h} D^{NPQ} -3\delta A^{(N}\wedge
A^{P} \wedge (F^{Q)}-G^{Q)}) +6 \Lambda^{(N}F^{P}\wedge
(F^{Q)}-G^{Q)})
 \nonumber  \\
& & \nonumber \\
& & -3 \Lambda^{(N}  (F^{P}-G^{P})\wedge  (F^{Q)}-G^{Q)})\, .
\end{eqnarray}





\begin{thebibliography}{99}










\bibitem{Bergshoeff:2001pv}
E.~Bergshoeff, R.~Kallosh, T.~Ort\'{\i}n, D.~Roest and A.~Van Proeyen,
``New formulations of D = 10 supersymmetry and D8 - O8 domain walls,''
Class.\ Quant.\ Grav.\  {\bf 18} (2001) 3359
[\hepth{0103233}].


\bibitem{Cremmer:1998px}
E.~Cremmer, B.~Julia, H.~Lu and C.~N.~Pope,
``Dualisation of dualities. II: Twisted self-duality of doubled fields  and
superdualities,''
Nucl.\ Phys.\  B {\bf 535} (1998) 242
[\hepth{9806106}].




\bibitem{Lavrinenko:1999xi}
I.~V.~Lavrinenko, H.~Lu, C.~N.~Pope and K.~S.~Stelle,
``Superdualities, brane tensions and massive IIA/IIB duality,''
Nucl.\ Phys.\  B {\bf 555} (1999) 201
[\hepth{9903057}].

\bibitem{West:2001as}
P.~C.~West,
``$E_{11}$ and M theory,
Class.\ Quant.\ Grav.\  {\bf 18} (2001) 4443
[\hepth{0104081}].

\bibitem{Schnakenburg:2001ya}
I.~Schnakenburg and P.~C.~West,
``Kac-Moody symmetries of IIB supergravity,
Phys.\ Lett.\  B {\bf 517} (2001) 421
[\hepth{0107181}].

\bibitem{Kleinschmidt:2003mf}
A.~Kleinschmidt, I.~Schnakenburg and P.~West,
``Very-extended Kac--Moody algebras and their interpretation at low  levels,
Class.\ Quant.\ Grav.\  {\bf 21} (2004) 2493
[\hepth{0309198}].

\bibitem{Riccioni:2007ni}
F.~Riccioni and P.~C.~West,
``E(11)-extended spacetime and gauged supergravities,''
JHEP {\bf 0802} (2008) 039
[\arxiv{0712.1795 [hep-th]}].

\bibitem{Nicolai:1987kz}
H.~Nicolai,
``The integrability of N=16 supergravity,''
Phys.\ Lett.\  B {\bf 194} (1987) 402.

\bibitem{Nicolai:2000sc}
H.~Nicolai and H.~Samtleben,
``Maximal gauged supergravity in three dimensions,''
Phys.\ Rev.\ Lett.\  {\bf 86} (2001) 1686
[\hepth{0010076}].

\bibitem{Nicolai:2001sv}
H.~Nicolai and H.~Samtleben,
``Compact and noncompact gauged maximal supergravities in three dimensions,''
JHEP {\bf 0104} (2001) 022
[\hepth{0103032}].

\bibitem{deWit:2005ub}
B.~de Wit, H.~Samtleben and M.~Trigiante,
``Magnetic charges in local field theory,''
JHEP {\bf 0509} (2005) 016
[\arxiv{hep-th/0507289}].

\bibitem{deWit:2008ta}
B.~de Wit, H.~Nicolai and H.~Samtleben,
``Gauged Supergravities, Tensor Hierarchies, and M-Theory,''
JHEP {\bf 0802} (2008) 044
[\arxiv{0801.1294 [hep-th]}].

\bibitem{deWit:2008gc}
B.~de Wit and H.~Samtleben,
``The end of the p-form hierarchy,''
JHEP {\bf 0808} (2008) 015
[\arxiv{0805.4767 [hep-th]}].

\bibitem{deWit:2007mt}
  B.~de Wit, H.~Samtleben and M.~Trigiante,
  ``The maximal D = 4 supergravities,''
  JHEP {\bf 0706} (2007) 049
  [arXiv:0705.2101 [hep-th]].

\bibitem{Vroome:2007zd}
 M.~de Vroome and B.~de Wit,
``Lagrangians with electric and magnetic charges of N=2 supersymmetric gauge
theories,''
JHEP {\bf 0708} (2007) 064
[\arxiv{0707.2717 [hep-th]}].


\bibitem{Bergshoeff:2007ij}
E.~A.~Bergshoeff, J.~Hartong, M.~H\"ubscher and T.~Ort\'{\i}n,
``Stringy cosmic strings in matter coupled N=2, d=4 supergravity,''
JHEP {\bf 0805} (2008) 033
[\arxiv{0711.0857} [hep-th]].

\bibitem{Bergshoeff:2007ef}
E.~Bergshoeff, H.~Samtleben and E.~Sezgin,
``The Gaugings of Maximal D=6 Supergravity,''
JHEP {\bf 0803} (2008) 068
[arXiv:0712.4277 [hep-th]].

\bibitem{Bergshoeff:2007vb}
E.~A.~Bergshoeff, J.~Gomis, T.~A.~Nutma and D.~Roest,
``Kac-Moody Spectrum of (Half-)Maximal Supergravities,''
JHEP {\bf 0802} (2008) 069
[\arxiv{0711.2035 [hep-th]}].

\bibitem{Hartong:2009az}
J.~Hartong, M.~H\"ubscher and T.~Ort\'{\i}n,
\arxiv{0903.0509} [hep-th].

\bibitem{Weidner:2006rp}
M.~Weidner,
``Gauged Supergravities in Various Spacetime Dimensions,''
Fortsch.\ Phys.\  {\bf 55} (2007) 843
[\hepth{0702084}].

\bibitem{Bergshoeff:2008qd}
E.~A.~Bergshoeff, O.~Hohm and T.~A.~Nutma,
``A Note on E11 and Three-dimensional Gauged Supergravity,''
JHEP {\bf 0805} (2008) 081
[\arxiv{0803.2989 [hep-th]}].

\bibitem{Gaillard:1981rj}
M.~K.~Gaillard and B.~Zumino,
``Duality Rotations For Interacting Fields,''
Nucl.\ Phys.\  B {\bf 193} (1981) 221.


\bibitem{Nicolai:2001ac}
H.~Nicolai and H.~Samtleben,
``N = 8 matter coupled AdS(3) supergravities,''
Phys.\ Lett.\  B {\bf 514} (2001) 165
[\hepth{0106153}].

\bibitem{deWit:2003ja}
B.~de Wit, I.~Herger and H.~Samtleben,
``Gauged locally supersymmetric D = 3 nonlinear sigma models,''
Nucl.\ Phys.\  B {\bf 671} (2003) 175
[\hepth{0307006}].



\bibitem{deWit:2004nw}
B.~de Wit, H.~Samtleben and M.~Trigiante,
``The maximal D = 5 supergravities,''
Nucl.\ Phys.\  B {\bf 716} (2005) 215
[\hepth{0412173}].


\bibitem{deWit:2002vt}
B.~de Wit, H.~Samtleben and M.~Trigiante,
``On Lagrangians and gaugings of maximal supergravities,''
Nucl.\ Phys.\  B {\bf 655} (2003) 93
[\hepth{0212239}].


\bibitem{Bergshoeff:2008cz}
 E.~A.~Bergshoeff, M.~de Roo and O.~Hohm,
``Multiple M2-branes and the Embedding Tensor,''
Class.\ Quant.\ Grav.\  {\bf 25} (2008) 142001
[\arxiv{0804.2201 [hep-th]}].

\bibitem{Boulanger:2008nd}
N.~Boulanger and O.~Hohm,
``Non-linear parent action and dual gravity,''
Phys.\ Rev.\  D {\bf 78} (2008) 064027
[\arxiv{0806.2775 [hep-th]}].

\bibitem{Bergshoeff:2008vc}
E.~A.~Bergshoeff, M.~de Roo, S.~F.~Kerstan, A.~Kleinschmidt and F.~Riccioni,
``Dual Gravity and Matter,''
\arxiv{0803.1963 [hep-th]}.

\bibitem{DeRydt:2008hw}
J.~De Rydt, T.~T.~Schmidt, M.~Trigiante, A.~Van Proeyen and M.~Zagermann,
``Electric/magnetic duality for chiral gauge theories with anomaly
cancellation,''
\arxiv{0808.2130 [hep-th]}.

\bibitem{BdeWit}
B.~de Wit and M.~van Zalk, ``Supergravity and M-theory'', Gen.\
Rel.\ Grav.\  {\bf 41} (2009) 757, proceedings of {\it Quantum
Gravity: Challenges and Perspectives}, Bad Honnef (Germany), April
2008. [arXiv:0901.4519 [hep-th]].















\end{thebibliography}
\end{document}